\documentclass[Afour,sagev,times]{sagej}
\usepackage{moreverb,url}

\usepackage[colorlinks,bookmarksopen,bookmarksnumbered,citecolor=red,urlcolor=red]{hyperref}
\usepackage{float}
\usepackage{latexsym,ifthen,rotating,calc,textcase,booktabs,color,endnotes}
\usepackage{amsfonts,amssymb,amsbsy,amsmath,amsthm}
\usepackage{bm}
\usepackage[errorshow]{tracefnt}
\usepackage{epstopdf}
\usepackage{soul}
\soulregister\cite7 
\soulregister\citep7 
\soulregister\citet7 
\soulregister\ref7 
\soulregister\pageref7 

\newcommand\BibTeX{{\rmfamily B\kern-.05em \textsc{i\kern-.025em b}\kern-.08em
T\kern-.1667em\lower.7ex\hbox{E}\kern-.125emX}}

\def\volumeyear{2021}

\begin{document}

\runninghead{Shan, Xu, Zhang, Wang and Chen}

\title{Discrete Boltzmann modeling of detonation: based on the Shakhov model}

\author{Yiming Shan\affilnum{1},Aiguo Xu\affilnum{1,2,3}, Yudong Zhang\affilnum{4}, Lifeng Wang\affilnum{1,3} and Feng Chen\affilnum{5}}

\affiliation{\affilnum{1}Laboratory of Computational Physics, Institute of Applied Physics and Computational Mathematics, Beijing 100088, China\\
\affilnum{2}State Key Laboratory of Explosion Science and Technology, Beijing Institute of Technology, Beijing 100081, China\\
\affilnum{3}HEDPS,Center for Applied Physics and Technology, and College of Engineering, Peking University, Beijing 100871, China\\
\affilnum{4}School of Mechanics and Safety Engineering, Zhengzhou University, Zhengzhou 450001, China\\
\affilnum{5} Shan Dong Jiaotong University, Jinan 250357, China
}
\corrauth{Aiguo Xu}

\email{Xu\_Aiguo@iapcm.ac.cn}

\begin{abstract}
A Discrete Boltzmann Model(DBM) based on the Shakhov model for detonation is proposed. Compared with the DBM based on the Bhatnagar-Gross-Krook (BGK) model, the current model has a flexible Prandtl numbers and consequently can be applied to a much wider range of detonation phenomena. Besides the Hydrodynamic Non-Equilibrium (HNE) behaviors usually investigated by the Navier-Stokes model, the most relevant Thermodynamic Non-Equilibrium (TNE) effects can be probed by the current model.
The model is validated by some well-known benchmarks,and some steady and unsteady detonation processes are investigated. As for
 the von Neumann peak relative to the wave front, it is found that
 (i) (within the range of numerical experiments) the peak heights of pressure, density and flow velocity increase exponentially with the Prandtl number, the maximum stress increases parabolically with the Prandtl number, and the maximum heat flux decreases exponentially with the Prandtl number;
(ii) the peak heights of pressure, density, temperature and flow velocity and the maximum stress within the peak are parabolically increase with the Mach number, the maximum heat flux decreases exponentially with the Mach number.

\end{abstract}

\keywords{Discrete Boltzmann method, combustion, detonation, shock wave, non-equilibrium}

\maketitle

\section{Introduction}
As a common phenomenon, combustion, including detonation, extensively exists in nature, our daily life, and various industrial fields \cite{ren2014numerical,ren2019supersonic, Ju-Review,Yang2013Large,wang2020laminar,wang2021effects}. The research of combustion problems is of great significance for actual production problems. In the combustion process, fluid dynamics and chemical reactions are coupled and interact with each other, and this process spans multiple orders of magnitude of time and space. These phenomena which are on different scales influence each other and show complex flow characteristics. For detonation, as a special case of combustion phenomenon, previous research mainly relied on experiments and a bit of theoretical analysis\cite{Ju-Review,Wu2018energy-fuel,Ma2018Energy,Ma2019Analysis}. With the development of computer technology, the numerical simulation of detonation has made great achievements \cite{2006Direct,wang2020laminar,wang2021effects,ren2019supersonic,ren2014numerical,ni2021numerical,watanabe2020numerical,Ma2021Fuel,Wjp2012,Xu2021AAAS,Xu2015APS}.
Generally, the simulation of detonation system has three scales: microscopic scale, mesoscopic scale and macroscopic scale. For the microscopic scale, molecular dynamics (MD) is a common method\cite{2016Molecular,2017Molecular}. This method can establish the reaction rate equation and provide complete information about the flow field. However, due to the large amount of calculation, the applicable space-time scale is restricted. The macroscopic scale usually refers to the Euler or Navier-Stokes (NS) equations with a phenomenological model describing the chemical reaction process. Traditional computational fluid dynamics (CFD) has made a great contribution to the research of detonation. But now, people have dealt with more and more complex situations, such as supersonic flow characterized by dramatic changes in the flow field. Especially in these extreme conditions, compressibility, nonlinearities, the discrete nature and strong coupling, as well as, other effects are important characteristics. And the rationality of the traditional fluid modeling theory is facing huge challenges. However, many of the above-mentioned problems occur in the time and space scales that cannot be simulated by molecular dynamics due to the calculation amount.

As for mesoscopic scale, it is mostly based on the Boltzmann equation. According to the kinetic theory and the Chapman-Enskog multi-scale analysis, the Euler equation assumes that the system is always thermodynamic equilibrium, and NS equation only contains the first-order non-equilibrium effects. However, complex phenomenon such as detonation have a large number of mesoscopic structures and kinetic models in general, which are weak in the related researches due to immature models and methods. As a mesoscopic method developed quickly in recent years, the discrete Boltzmann method (DBM) has been applied in many fields\cite{Xu-Chapter2,Xu2021AAS,Xu2021CJCP,Ji2021AIPA,Lin2020Entropy,Lin2018CAF,Lin2018CNF,ChenL2021FOP,Xu2018FOP}. It can be regarded as a variant hybrid of the Lattice Boltzmann Method (LBM)\cite{Succi2001book,shan1993lattice,zhang2005lattice,ambrus2019quadrature,chen2010multiple,li2012additional,wang2020simple,chen2018highly,wang2020simplified,saadat2020semi,
fei2019modeling,qiu2020study,qiu2021mesoscopic,sun2020discrete,sun2019anisotropic,zhan2021lattice,Huang2021transition,Wang2021Lattice,Liu2020CNF-LBM,2015Improvement,2014A,ghadyani2015use} and the description method of non-equilibrium behavior in statistical physics. In the absence of misunderstanding, DBM is also used as an abbreviation for discrete Boltzmann model or discrete Boltzmann modeling.

Historically, there are in fact two branches in the LBM studies. One branch regards the LBM as a novel numerical scheme to solve Partial Differential Equation(s)(PDE). The other branch regards the LBM as a novel mesoscopic kinetic modeling method. The latter works as a more fundamental-level physical modeling approach for fluid systems, as opposed to macro modeling.
Since the goals of the two branches are different, the rules of construction for them are different. In the past three decades, the LBM for solving PDF achieved great success in various fields. In contrast, before 2012, the LBM modeling method did not present noticeable difference from the LBM for solving PDE, except for more strictly follow the physical connotations of Boltzmann equation and kinetic moments. Consequently, the LBM in literature gradually became a synonym for LBM for solving PDE. In 2012 Xu, et al. \cite{Xu2012PoF} proposed to use the non-conservative kinetic moments of $(f - f^{eq})$ to describe how and how much the system deviates from its thermodynamic state and the corresponding thermodynamic non-equilibrium effects, where $f$ and $f^{eq}$ are the distribution function and corresponding equilibrium distribution function, respectively. This is the starting point of the LBM modeling method found its physical function beyond the Navier-Stokes model. Later, Xu, et al. \cite{1995Multiple} further proposed to use the phase space opened by the non-conservative kinetic moments of $(f - f^{eq})$ to describe the system state and evolution. Up to this point, the complex thermodynamic non-equilibrium behaviors found an intuitive geometrical correspondence.
It is clear that DBM was developed from the branch of LBM which works as a building method of physical model for flow system.

Compared with the traditional computational fluid dynamics methods, DBM is a basic and underlying theoretical model construction method, which is to observe and describe the physical system from a broader viewpoint. Besides the conserved moments of the distribution function (conservation of mass, momentum, and energy), DBM also pays attention to the temporal and spatial evolution of non-conserved moments that are related to our researches. And more importantly, DBM provides a set of complex physical field analysis methods. In DBM, we can use the difference of non-conserved moments of distribution function and local equilibrium distribution function to describe the non-equilibrium behavior of complex flows. Each of the independent components of these non-conserved moments describes the degree of system deviating from thermal equilibrium from different perspectives, and it is also the manifestation of the physical effects caused by the flow system deviating from the thermal equilibrium. Some new non-equilibrium information that is hard to describe previously can be stratified and quantitatively studied by using ``non-equilibrium intensity" and related concepts.

Extensive efforts have been devoted to the studies of the detonation phenomenon. Watanabe \emph{et al}.\cite{watanabe2020numerical} researched the mean structure of gaseous detonation laden with a dilute water spray by the two-dimensional Eulerian - Lagrangian method and obtained relevant results. Ni \emph{et al}.\cite{ni2021numerical} used a symmetric two-phase detonation model and a cylinder heat conduction model to research the features of transient heat transfer in pulse detonation engine (PDE), and the relevant numerical results are consistent with the experimental results, and this research can provide help for the PDEs design. Kolera-Gokula \emph{et al}. \cite{2006Direct} studied the unsteady interaction between a vortex pair and a premixed flame kernel by using direct numerical simulations and they proposed a new parameter which can evaluate the fraction of mutually interacting flames. As for DBM's progress in detonation research, Zhang \emph{et al}. \cite{Zhang2016CNF} deduced a new set of hydrodynamic equations which viscous stress tensor and heat flux are replaced by two non-equilibrium quantities, and studied four kinds of detonation phenomena with different reaction rates in which Negative Temperature Coefficient (NTC) regime was included and non-equilibrium quantities and entropy productions were researched. Lin \emph{et al}. \cite{Lin2016CNF} proposed a double-distribution-function discrete Boltzmann model for detonation (combustion), and verified that this model can be applied to the simulations of subsonic and supersonic combustion phenomena. As we all know, the BGK model has been widely used, but there is an obvious defect: the Prandtl number ($\rm{Pr}$) is fixed in $1$.\cite{Zhang2019CPC,Zhang2018FOP} (In BGK models, the viscosity coefficient $\mu {\rm{ = }}\tau P$, where $\tau $ is relaxation time, and $P$ is pressure. Thermal conduction coefficient $\kappa {\rm{ = }}{C_p}\tau P$ , where ${C_p}$ is the specific heat at constant pressure. And $\rm{Pr}$ is defined as: $\Pr  = \frac{{{C_p}\mu }}{\kappa } = \frac{{{C_p}\tau P}}{{{C_p}\tau P}} = 1$.
) At present, there are mainly two methods to overcome this defect: (i) using the Multiple Relaxation Time (MRT) model\cite{1995Multiple,Lin2021PRE,Xu2015APS,Chen2016FOP,Chen2018POF}, and (ii) reasonably modifying the framework of the Single Relaxation Time (SRT) BGK-like model \cite{Zhang2019CPC,Zhang2018FOP,Zhang2020PoF-2Fluid}, and both of them have their own advantages and strengths. At present, in the DBM modeling including chemical reactions, the previous works to solve the problem of fixed $\rm{Pr}$ numbers are mainly based on the first idea\cite{1995Multiple,Lin2021PRE}. In this paper, we propose a discrete Boltzmann model for detonation based on the Shakhov model, so that the $\rm{Pr}$ number can be adjusted in the SRT framework. The model is validated by some well-known benchmarks, and some unsteady detonation processes are investigated.

This paper is organized as follows. In section II, the DBM and the physical quantities used to describe the combustion system are briefly introduced. Systematic numerical simulations and analyses are shown in Section III. A brief conclusion is given in Section IV.

\section{Simulation methods}
\subsection{Shakhov model}
The Shakhov model was proposed by Shakhov in 1968\cite{Shakhov1968}, and this model modify the local equilibrium distribution function $f^{eq}$ to implement adjustable $\rm{Pr}$, and this new distribution function is called Shakhov distribution function $f^{S}$,
\begin{equation}\label{Eq:fs}
\begin{aligned}
& {f^S} ={f^{eq}} +{f^{eq}}\\
& \left\{ {(1 - \Pr ){c_\alpha }{q_\alpha }[\frac{{{c^2} + {\eta ^2}}}{{RT}} - (D + n + 2)]/[(D + n + 2)PRT]} \right\} ,
\end{aligned}
\end{equation}
where
\begin{equation}\label{Eq:feq}
{f^{eq}} = \rho {\left( {\frac{1}{{2\pi RT}}} \right)^{(D + n)/2}}\exp \left( { - \frac{{{c^2} + {\eta ^2}}}{{2RT}}} \right),
\end{equation}
$f^{eq}$ indicates the Maxwell distribution function, and $\rm{Pr}$ is Prandtl number. ${c_\alpha }{\rm{ = }}{v_\alpha } - {u_\alpha }$ is the microscopic fluctuation of the molecular velocity in the $\alpha $ direction, $\alpha $ can be $x$,$y$ or $z$. ${u_\alpha }$ is macroscopic velocity, ${c^2} = {c_\chi}{c_\chi}$, $D$ is the spatial dimension, $n$ is the number of extra degrees of freedom, $R$ is the ideal gas constant, $P$ is macroscopic pressure, and $T$ is temperature. ${q_\alpha }$ is the heat flux in $\alpha $ direction, ${\eta ^2}$ is the total energy of the extra degree of freedom. It can be seen from the above equation that Shakhov distribution function ${f^S}$ modifies the heat flux term in the ${f^{eq}}$ in fact.

\subsection{The DBM model}

Based on the literatures \cite{Xu2015APS,Zhang2019CPC}, the chemical reaction contribution is added to collision term of the Boltzmann equation.
 As a result, the evolution equation becomes,
\begin{equation}\label{Eq:DBM}
\frac{\partial }{{\partial t}}\left( {\begin{array}{*{20}{c}}
g\\
h
\end{array}} \right) + {v_\alpha }\frac{\partial }{{\partial {r_\alpha }}}\left( {\begin{array}{*{20}{c}}
g\\
h
\end{array}} \right) =  - \frac{1}{\tau }\left( {\begin{array}{*{20}{c}}
{g - {g^S}}\\
{h - {h^S}}
\end{array}} \right) + C ,
\end{equation}
where $g = \int {fd} {\bf{\eta }}$, $h = \int {f\frac{{{\eta ^2}}}{2}} d{\bf{\eta }}$, $f$ is the distribution function, the purpose of this processes are eliminate the dependence of distribution function $f$ on extra degrees of freedom, ${v_\alpha }$ is the molecular velocity in the $\alpha $ direction. ${r_\alpha }$ is the space variable, $t$ is time, $\tau $ is the relaxation time. ${g^S}$  and ${h^S}$ are the distribution functions in the Shakhov model \cite{Zhang2019CPC},
\begin{equation}\label{Eq:DBM-gs}
\begin{aligned}
& {g^S} = {g^{eq}} + {\kern 1pt} {\kern 1pt} {g^{eq}}\\
& {\kern 1pt} {\kern 1pt} {\kern 1pt} \left\{ {\left( {1 - \Pr } \right){c_\alpha }{q_\alpha }\left[ {\frac{{{c^2}}}{{RT}} - \left( {D + 2} \right)} \right]/\left[ {\left( {D + n + 2} \right)PRT} \right]} \right\},
\end{aligned}
\end{equation}
\begin{equation}\label{Eq:DBM-hs}
\begin{aligned}
& {h^S} = {h^{eq}} + \\
& {\kern 1pt} {\kern 1pt} {\kern 1pt} {\kern 1pt} {\kern 1pt} {\kern 1pt} {h^{eq}}\left\{ {\left( {1 - \Pr } \right){c_\alpha }{q_\alpha }\left( {\frac{{{c^2}}}{{RT}} - D} \right)/\left[ {\left( {D + n + 2} \right)PRT} \right]} \right\}
\end{aligned} ,
\end{equation}
where
\begin{equation}\label{Eq:DBM-geq}
{g^{eq}} = \rho {\left( {\frac{1}{{2\pi RT}}} \right)^{D/2}}\exp \left( { - \frac{{{c^2}}}{{2RT}}} \right) ,
\end{equation}
\begin{equation}\label{Eq:DBM-heq}
{h^{eq}} = \frac{{nRT}}{2}{g^{eq}}.
\end{equation}
It can be found that the dependence of the velocity distribution function on the extra degree of freedom is eliminated by the above treatment, and the expressions of ${g^{eq}}$ and ${g^S}$  do not contain extra degree of freedom. Besides, ${h^{eq}}$ and ${h^S}$  can be obtained by  ${g^{eq}}$ and ${g^S}$ conveniently. $C$ is the chemical reaction term, and the following assumptions are made here:\\
i) There are only two kinds of substances in the detonation process: reactant and product. The flow field can be described by one distribution function $f$, and the relaxation time $\tau $ does not change with density $\rho $ and temperature $T$;\\
ii) Heat radiation effect is ignored;\\
iii) The time scale of chemical reactions is much larger than the time scale of thermodynamic relaxation time $\tau $;\\
iv) The chemical reaction process is irreversible.\\
Besides, the reaction process can be described by the following reaction rate equation,
\begin{equation}\label{Eq:DBM-2}
F(\lambda ) = \frac{{d\lambda }}{{dt}} = {\lambda ^\prime },
\end{equation}
where $\lambda $ is the reaction process parameter, $F\left( \lambda  \right)$ is the reaction rate function. The chemical reaction rate model used in this article is\cite{Zhang2016CNF}
\begin{equation}\label{Eq:DBM-21}
\frac{{d\lambda }}{{dt}} = \left\{ {\begin{array}{*{20}{l}}
{k(1 - \lambda )\lambda {\kern 1pt} {\kern 1pt} {\kern 1pt} {\kern 1pt} {\kern 1pt} {\kern 1pt} ,{\kern 1pt} {\kern 1pt} {\kern 1pt} {\kern 1pt} {\kern 1pt} {\kern 1pt} T \ge {T_{th}}{\rm{ }}{\kern 1pt} {\kern 1pt} {\kern 1pt} {\rm{and }}{\kern 1pt} {\kern 1pt} {\kern 1pt} 0 \le \lambda  \le 1}\\
{0,{\rm{            }}{\kern 1pt} {\kern 1pt} {\kern 1pt} {\kern 1pt} {\kern 1pt} {\kern 1pt} {\kern 1pt} {\kern 1pt} {\kern 1pt} {\kern 1pt} {\kern 1pt} {\kern 1pt} {\kern 1pt} {\kern 1pt} {\kern 1pt} {\kern 1pt} {\kern 1pt} {\kern 1pt} {\kern 1pt} {\kern 1pt} {\kern 1pt} {\kern 1pt} {\kern 1pt} {\kern 1pt} {\kern 1pt} {\kern 1pt} {\kern 1pt} {\kern 1pt} {\kern 1pt} {\kern 1pt} {\kern 1pt} {\kern 1pt} {\kern 1pt} {\kern 1pt} {\kern 1pt} {\kern 1pt} {\kern 1pt} {\kern 1pt} {\kern 1pt} {\kern 1pt} {\kern 1pt} {\kern 1pt} {\kern 1pt} {\kern 1pt} {\kern 1pt} {\kern 1pt} {\kern 1pt} {\kern 1pt} {\kern 1pt} {\rm{ }}else}
\end{array}} \right. ,
\end{equation}
due to the purpose of this paper is to research non-equilibrium behaviors in the detonation processes, the chemical reaction rate model is a simplified model. Where $k$ is reaction rate coefficient, and ${T_{th}}$ is the blasting temperature.

And then, we can obtain the expression of the chemical reaction term,
\begin{equation}\label{Eq:DBM-3}
C = \frac{1}{\tau }\left( {\begin{array}{*{20}{c}}
{{g^{*S}} - {g^S}}\\
{{h^{*S}} - {h^S}}
\end{array}} \right),
\end{equation}
where ${g^{*S}}$ and ${h^{*S}}$ are the local equilibrium distribution function after adding the chemical reaction in Shakhov model, and
\begin{equation}\label{Eq:DBM-4}
{g^{*S}} = {g^S}(\rho ,{\bf{u}},e + \tau \rho QF(\lambda )),
\end{equation}
\begin{equation}\label{Eq:DBM-5}
{h^{*S}} = {h^S}(\rho ,{\bf{u}},e + \tau \rho QF(\lambda )),
\end{equation}
where $Q$ is the heat released from the unit mass reactant that react completely, and $e$ is the internal energy.

\begin{figure}
\centering
\includegraphics[height=7cm]{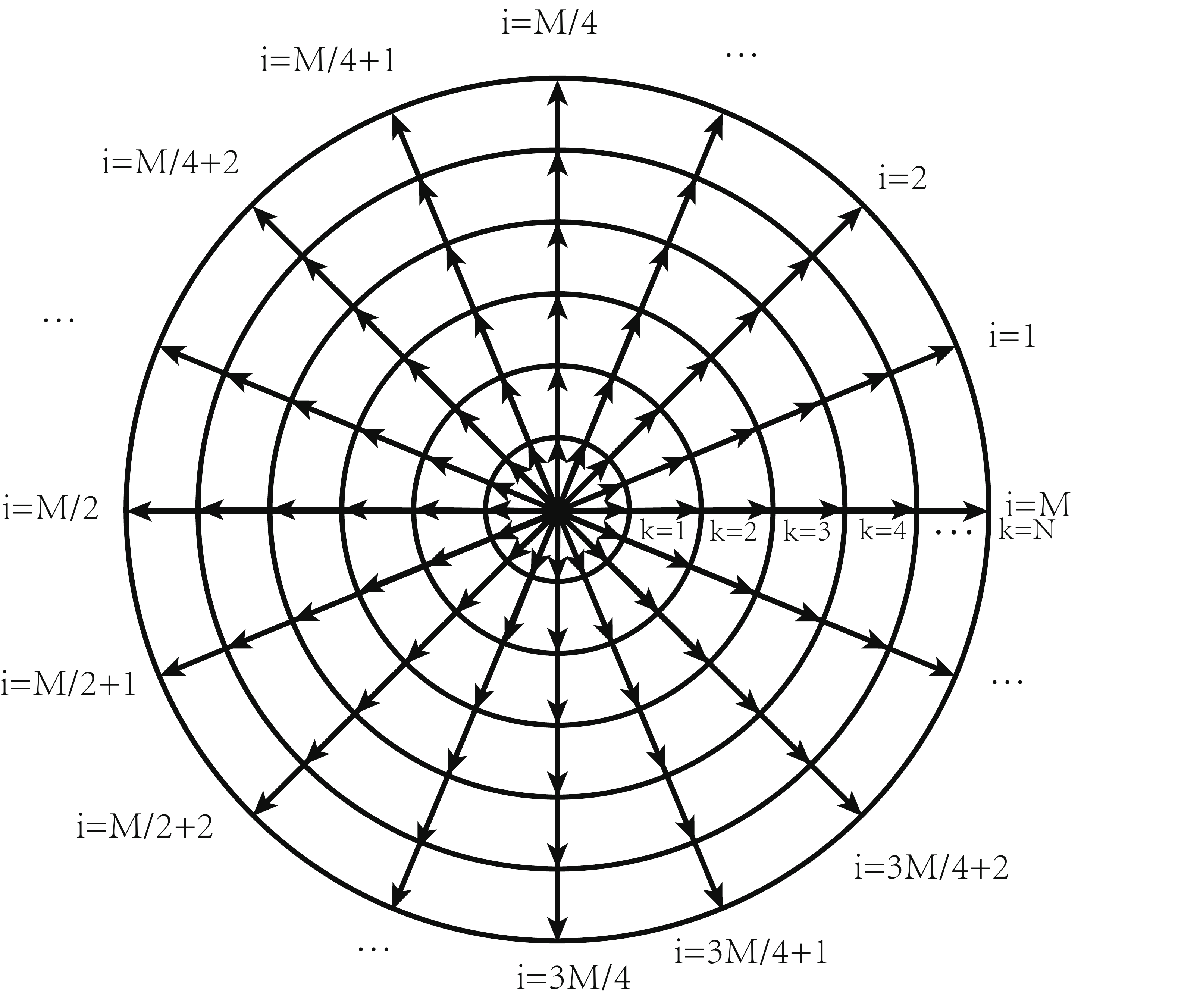}
\caption{Schematic of discrete velocity model.}
\label{Fig:DVM1}
\end{figure}

The basic principle of DBM discretization is that the kinetic moments of concern must remain values unchanged after being converted into sums for calculation. It should be pointed out that DBM gives physical constraints about discrete speeds only, and does not involve any specific discretization formats. The discrete velocity model selected in this paper is shown in the Fig.\ref{Fig:DVM1}, and
\begin{equation}\label{Eq:DBM-6}
{{\bf{v}}_{ki\alpha }} = {v_k}{e_{i\alpha }} ,
\end{equation}
where $k = 0,1, \cdots ,N $ , that is, the discrete velocity used in this research includes a zero velocity and $N $ non-zero velocity. Except for the zero velocity, the value of non-zero discrete velocity in each group is the same, but the direction is different. ${e_{i\alpha }}$ is a unit vector to represent the direction of the discrete velocity ${v_{ki\alpha }}$ , and for two - dimensional cases, ${e_{i\alpha }} $ is described by the following equation,
\begin{equation}\label{Eq:DBM-7}
\left( {{e_{ix}},{e_{iy}}} \right) = \left[ {\cos \left( {\frac{{i - 1}}{M} \cdot 2\pi } \right),\sin \left( {\frac{{i - 1}}{M} \cdot 2\pi } \right)} \right]
\end{equation}
$$i = 1,2, \cdots ,M ,$$
where $M$ is the number of each group of discrete velocity, and $M$ should satisfy the following relationship,
\begin{equation}\label{Eq:DBM-8}
M \ge 2N + 1 .
\end{equation}

So, the evolution equation of DBM with chemical reaction reads
\begin{equation}\label{Eq:DBM-shakhov}
\begin{aligned}
& \frac{\partial }{{\partial t}}\left( {\begin{array}{*{20}{c}}
{{g_{ki}}}\\
{{h_{ki}}}
\end{array}} \right) + {v_{ki\alpha }}\frac{\partial }{{\partial {r_\alpha }}}\left( {\begin{array}{*{20}{c}}
{{g_{ki}}}\\
{{h_{ki}}}
\end{array}} \right) =\\
& - \frac{1}{\tau }\left( {\begin{array}{*{20}{c}}
{{g_{ki}} - g_{ki}^S}\\
{{h_{ki}} - h_{ki}^S}
\end{array}} \right) + {C_{ki}}
,
\end{aligned}
\end{equation}
where $ {g_{ki}} $ and $ {h_{ki}} $ are the discrete distribution functions of $ g $ and $ h $, respectively, and $ {C_{ki}} $  is the chemical reaction term.  and
\begin{equation}\label{Eq:DBM-9}
g_{ki}^{eq} = \rho {F_k}\sum\limits_{n = 0}^N {\frac{1}{n}} \sum\limits_{{\eta _1}{\eta _2} \cdots {\eta _n}} {{\bf{H}}_{{\eta _1}{\eta _2} \cdots {\eta _n}}^{(n)}{T^{ - n/2}}{u_{{\eta _1}}}} {u_{{\eta _2}}} \cdots {u_{{\eta _n}}} ,
\end{equation}
where $ {F_k} $ is the weight coefficient,  $ {\bf{H}}_{{\eta _1}{\eta _2} \cdots {\eta _n}}^{(n)} $ is the $ n $ order Hermite polynomial about $ {v_{ki\eta }} $ , and the expression of  $ {\bf{H}}_{{\eta _1}{\eta _2} \cdots {\eta _n}}^{(n)} $  can be found in the literature \cite{Zhang2019CPC}. Weight coefficient ${F_k}$ can be calculated by the following equations,
\begin{equation}\label{Eq:DBM-10}
{F_k} = \frac{{\sum\limits_{n = 1}^N {{B_n}{G_{N - n}}\left( {{v_1},{v_2}, \cdots ,{v_{k - 1}},{v_{k + 1}}, \cdots ,{v_N}} \right)} }}{{v_k^2\prod _{n = 1,n \ne k}^N(v_k^2 - v_n^2)}} ,
\end{equation}
\begin{equation}\label{Eq:DBM-11}
{F_0}{\rm{ = }}1 - {B_0}\sum\limits_{n = 1}^N {{F_n}} ,
\end{equation}
where
\begin{equation}\label{Eq:DBM-12}
{B_n} = \left\{ {\begin{array}{*{20}{c}}
{M,}&{n = 0}\\
{{{( - 1)}^{n + N}}\frac{{2n!!}}{M}{T^n},}&{n \ne 0}
\end{array}} \right. ,
\end{equation}
\begin{equation}\label{Eq:DBM-13}
{G_{N - n}} = \left\{ {\begin{array}{*{20}{c}}
{1,}&{n = N}\\
{\sum\limits_{{m_1}  <  \cdots  < {m_{N - n}}}^{N - 1} {x_{{m_1}}^2 \cdots x_{{m_{N - n}}}^2,} }&{1 \le n \le N}
\end{array}} \right. .
\end{equation}
So, the expression of discrete local equilibrium distribution function $ g_{ki}^{eq} $ can be obtained , and
\begin{equation}\label{Eq:DBM-14}
h_{ki}^{eq} = \frac{{nT}}{2}g_{ki}^{eq} .
\end{equation}

Then, discrete distribution functions based on Shakhov model $ g_{ki}^S $ and $ h_{ki}^S $ can be obtained from the discrete local equilibrium distribution functions $ g_{ki}^{eq} $ and $ h_{ki}^{eq} $, respectively,
\begin{equation}\label{Eq:DBM-15}
\begin{aligned}
& g_{ki}^S = g_{ki}^{eq} + {\kern 1pt} {\kern 1pt} {\kern 1pt} g_{ki}^{eq}\\
& \left\{ {(1 - \Pr ){c_{ki\alpha }}{q_\alpha }\left[ {\frac{{c_{ki}^2}}{T} - \left( {D + 2} \right)} \right]/\left[ {\left( {D + n + 2} \right)PT} \right]} \right\} ,
\end{aligned}
\end{equation}
\begin{equation}\label{Eq:DBM-16}
\begin{aligned}
& h_{ki}^S = h_{ki}^{eq} +{\kern 1pt} {\kern 1pt} {\kern 1pt} h_{ki}^{eq}\\
& \left\{ {(1 - \Pr ){c_{ki\alpha }}{q_\alpha }\left( {\frac{{c_{ki}^2}}{T} - D} \right)/\left[ {\left( {D + n + 2} \right)PT} \right]} \right\} ,
\end{aligned}
\end{equation}
where
\begin{equation}\label{Eq:DBM-17}
{c_{ki\alpha }} = {v_{ki\alpha }} - {u_\alpha },
\end{equation}
\begin{equation}\label{Eq:DBM-18}
c_{ki}^2 = {c_{ki\beta }}{c_{ki\beta }}.
\end{equation}

The acquirement and analysis of non-equilibrium information in the flow field are the purpose and core of DBM. In DBM, the non-equilibrium behavior characteristics can be described in detail by the non-conserved kinetic moment of $ ({f_i} - f_i^{eq}) $ , and $ {{\bf{\Delta }}_n} $ is defined as the thermo-hydrodynamic non-equilibrium characteristic quantity,
\begin{equation}\label{Eq:DBM-19}
{{\bf{\Delta }}_n} = {{\bf{M}}_n}({f_i}) - {{\bf{M}}_n}(f_i^{eq}).
\end{equation}

And  $ {{\bf{\Delta }}_n} $ describes the details of the flow field deviating from equilibrium. Kinetic moment  $ {{\bf{M}}_n} $ and non-equilibrium characteristic quantity $ {{\bf{\Delta }}_n} $ contain the average behavior of fluid molecules and the purely thermal fluctuation behavior. In the same way, the $ {\bf{M}}_n^* $ is used to describe the $ n $ -order kinetic moment of the molecule speed fluctuation  $ \left( {{\bf{v}} - {\bf{u}}} \right) $ about the distribution function $ f $ , which is called central moment; And
$ {\bf{M}}_n^{*eq} = {\bf{M}}_n^*({f^{eq}}) $ ,
so the non-equilibrium characteristic quantity defined by the central moment reads
\begin{equation}\label{Eq:DBM-20}
{\bf{\Delta }}_n^* = {\bf{M}}_n^*({f_i}) - {\bf{M}}_n^*(f_i^{eq}),
\end{equation}
where ${\bf{\Delta }}_n^* $  is the characteristic quantity of thermodynamic non-equilibrium (TNE).

\begin{figure}
\centering
\includegraphics[height=10cm]{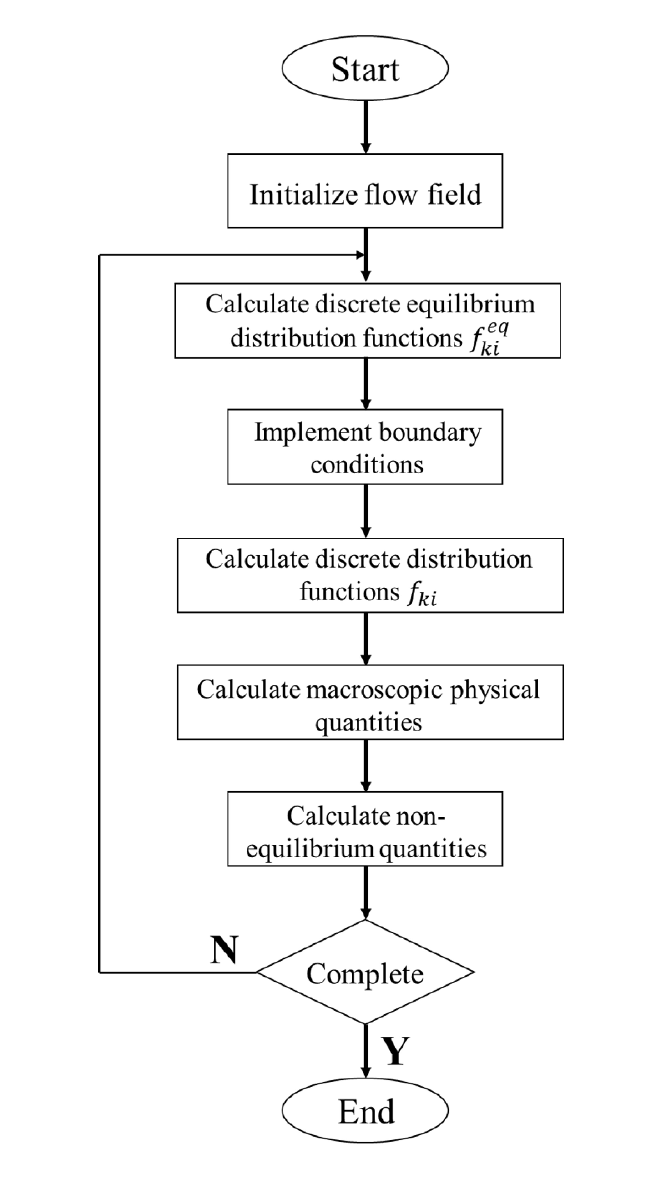}
\caption{Schematic algorithm for DBM simulation process.}
\label{Fig:DBM1}
\end{figure}

In summary, the schematic algorithm about DBM simulation process is shown in the Fig. \ref{Fig:DBM1}.

\section{Simulation results}

In order to verify the correctness and accuracy of the new model and research the non-equilibrium behavior, some numerical simulations have been carried out in this chapter. This chapter is divided into two parts: i) DBM simulation without chemical reaction, including Couette flow and Sod's shock tube problem. ii) DBM simulation with chemical reaction, including one-dimensional detonation wave propagation, detonation wave and shock wave collision and the two-dimensional detonation problem.

\subsection{Couette flow}

First, numerical simulations of some cases without the chemical reaction term are carried out. Couette flow is a classic case and is often used to verify the correctness of the model in CFD. The liquid is filled between the upper and lower plates and the distance between the two plates is  $ L $ .The liquid velocity is zero initially. The temperature of upper and lower plates are  $ {T_0} $ . Now, it is assumed that the upper plate moves in the positive direction of the $x$-axis at the speed  $ U $ , while the lower plate keeps still.
And then, the liquid will keep moving horizontally due to fluid viscous effect. Before reaching steady state ,the velocity of liquid in vertical coordinate $ y $ varies with time  $ t $  is as the following formula,
\begin{equation}\label{Eq:RES-1}
\begin{aligned}
& \frac{u}{U} = \frac{y}{L} - \frac{2}{\pi }\sum\limits_{n = 1}^\infty\\
& {\left[ {\frac{{{{( - 1)}^n}}}{n}\exp \left( { - {n^2}{\pi ^2}\frac{{\mu t}}{{\rho {L^2}}}} \right)\sin \left( {n\pi \left( {1 - \frac{y}{L}} \right)} \right)} \right]}.
\end{aligned}
\end{equation}
When the flow field is stable, the temperature of liquid $ T $ varies with $ y $ can usually be expressed as
\begin{equation}\label{Eq:RES-2}
T = {T_0} + \frac{{\Pr {U^2}}}{{2{C_p}L}}y(1 - \frac{y}{L}).
\end{equation}
This case is simulated, with the initial configuration,
\begin{equation}\label{Eq:RES-3}
\left( {\rho ,{u},{v},T} \right) = \left( {1.0,0,0,1.0} \right)
\end{equation}
where $\rho$ is density, $u$ is the velocity in the $x$ direction, $v$ is the velocity in the $y$ direction, $T$ is temperature. And other numerical simulation parameters are $U = 0.8$ , ${T_0} = 1.0 $, spatial and temporal steps are $ \Delta x = \Delta y = 0.002 $ and $ \Delta t = 1.0 \times {10^{ - 5}} $ ,respectively. Relaxation time is $ \tau {\rm{ = }}1 \times {10^{{\rm{ - }}3}} $ , and specific heat ratio is $ \gamma  = 7/5 $ . Number of grids in x-direction is $Nx=1$, and as for y-direction, the grid-independence verification of this case is implemented.

\begin{figure*}[ht]
\centering
\includegraphics[height=6.3cm]{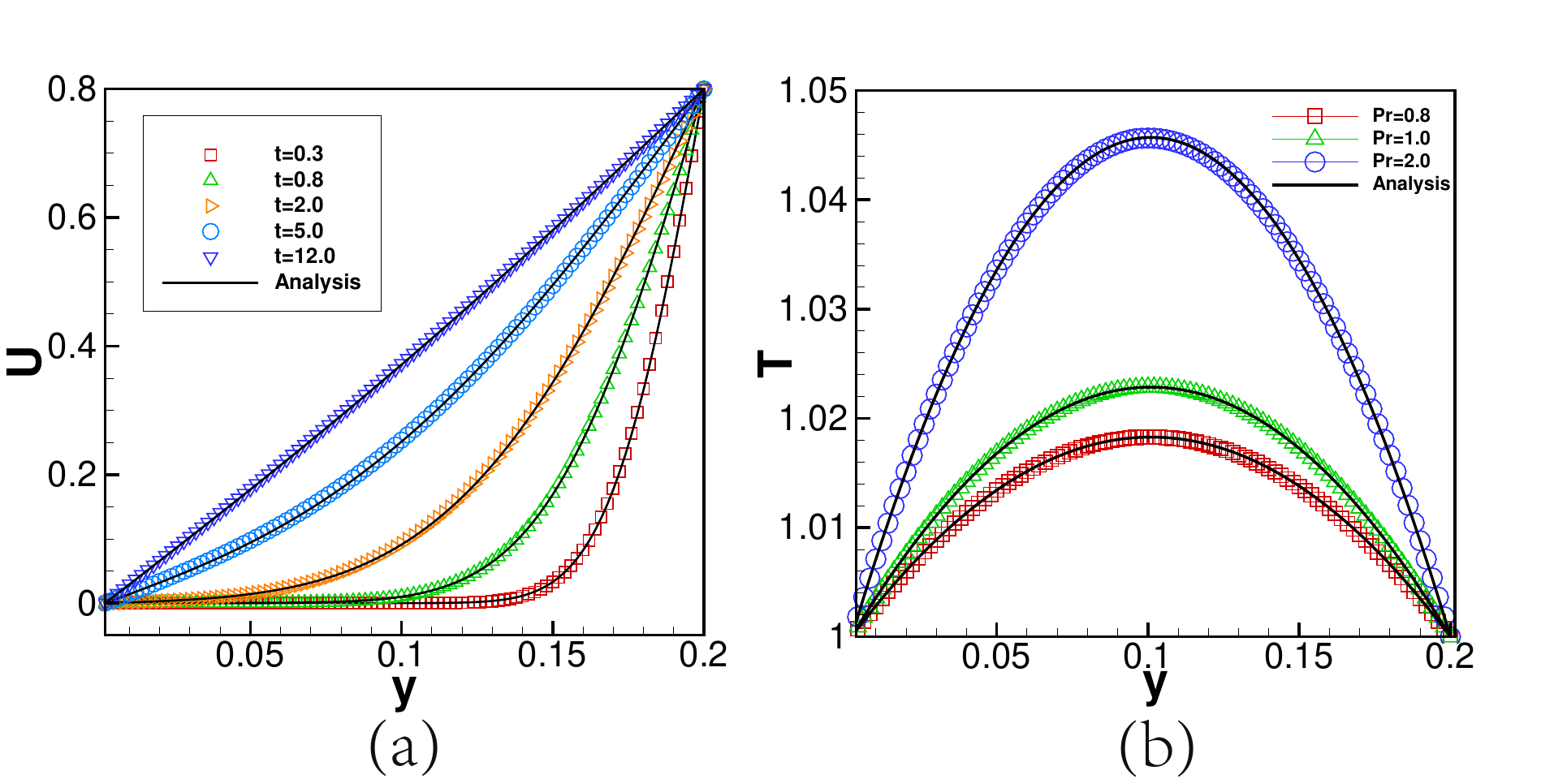}
\caption{Profiles of (a) velocity at $t=0.3$, $t=0.8$, $t=2$, $t=8.8$ and $t=12$, (b) temperature when fluid field is stable in $\rm{Pr}=0.8$, $\rm{Pr}=1.0$ and $\rm{Pr}=2.0$. }
\label{Fig:couette1}
\end{figure*}

\begin{figure*}[htbp]
\centering
\includegraphics[height=7cm]{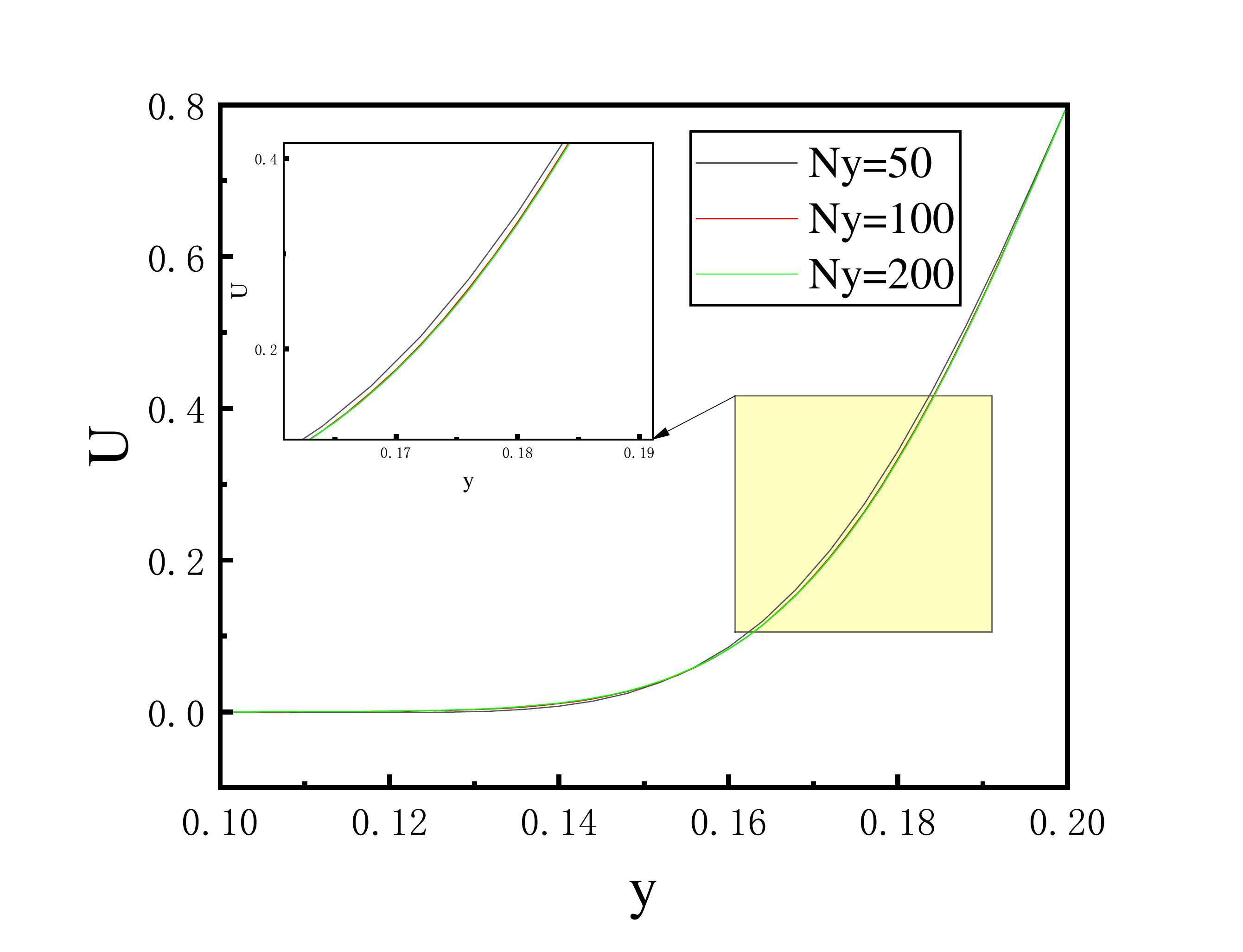}
\caption{Velocity $U$-$y$ diagram at $t=0.3$ with different $Ny$ . Simulation results have converged when $Ny$ is 100 (Red line)}
\label{Fig:NX1}
\end{figure*}

Velocity-$y$ diagram is selected at $t=0.3$, and the results of grid-independence verification are performed in Fig.\ref{Fig:NX1}, and the grid mesh of simulation are $ Ny=50, 100$ and $200$, respectively. The results have converged when the number of meshes $ Ny $ is 100, so the grid mesh is $ Nx \times Ny = 1 \times 100 $ in this case. Figure \ref{Fig:couette1} is the comparison of the DBM simulation results of couette flow with the analytical solution, and Fig.\ref{Fig:couette1}(a) is the velocity in the $ y $ at different time, and the solid line is analytical solution and the symbols is DBM simulation results. Figure \ref{Fig:couette1} (b) is the temperature of fluid in $\rm{Pr}=0.8, 1.0$ and $2.0$ when fluid field is stable and the L1-norms are 0.00171, 0.0027 and 0.00421, respectively. It is evident that the simulation results in different time and $\rm{Pr}$ agree well with the analytical solutions Eqs.\eqref{Eq:RES-1} and \eqref{Eq:RES-2}. Consequently, this case indicates the ability of the model to adjust the Prandtl number is verified.

\subsection{Sod $'$s shock tube}

A tube is divided into two parts: the left side and the right side, and the initially macroscopic physical quantities on the left and right sides are shown in the following formula:
\begin{equation}\label{Eq:RES-4}
{(\rho ,{u},{v},T)_L} = (1,0,0,1),
\end{equation}
\begin{equation}\label{Eq:RES-5}
{(\rho ,{u},{v},T)_R} = (0.125,0,0,0.8),
\end{equation}
where the subscript $ L $ and $ R $ represent the left part shock tube and the right, respectively.\\

The other simulation parameters are $ \Delta x = \Delta y = 1 \times {10^{ - 3}} $ , $ \Delta t = 2 \times {10^{ - 6}} $ , $ \tau {\rm{ = }}2 \times {10^{{\rm{ - }}4}} $ , and grid-independence verification of this case is also implemented.

\begin{figure*}[htbp]
\centering
\includegraphics[height=7cm]{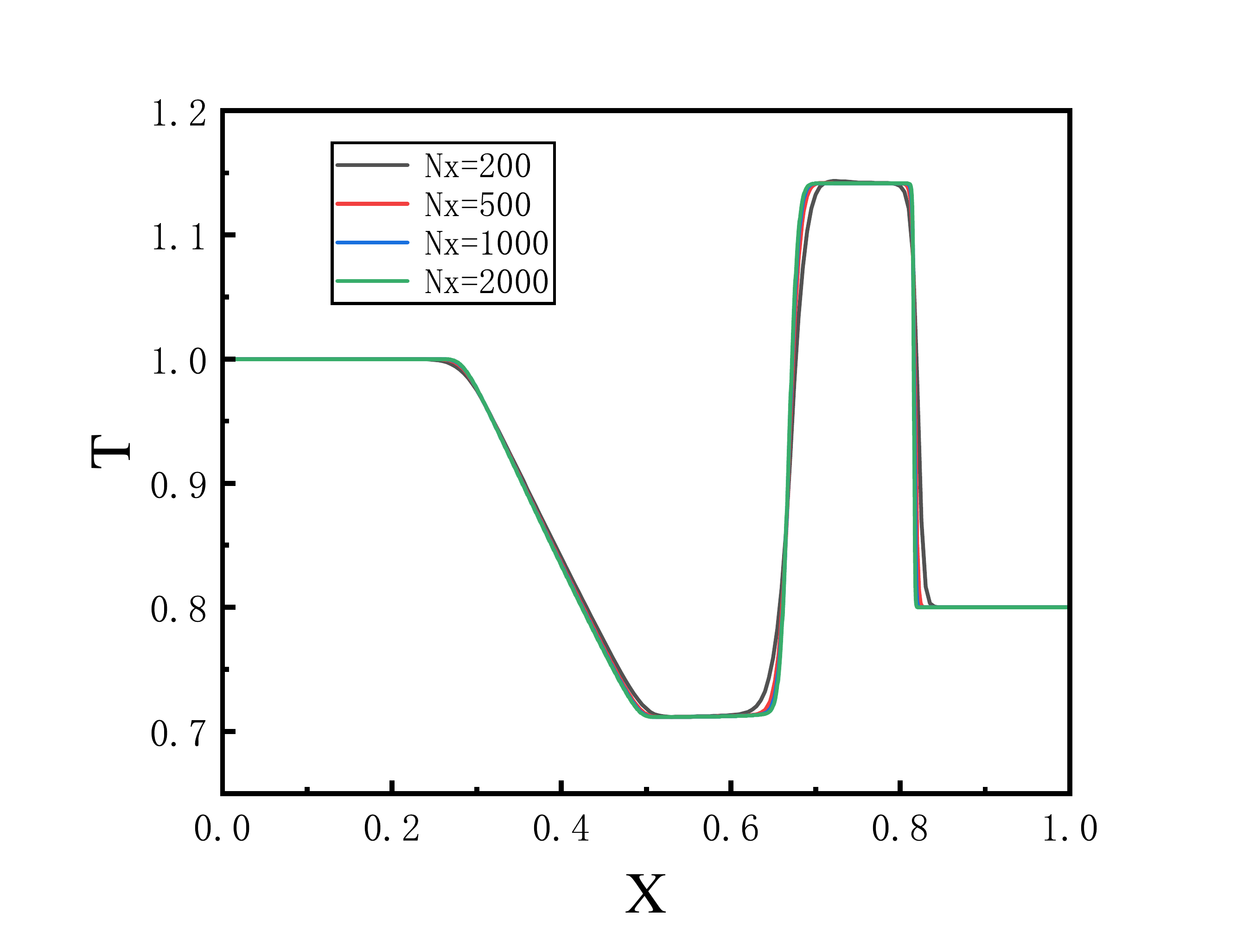}
\caption{Temperature $T$-$x$ diagram at $t = 0.18$ with different $Nx$ . Simulation results have converged when $Nx$ is 1000 (Blue line)}
\label{Fig:NX2}
\end{figure*}

\begin{figure*}[htbp]
\centering
\includegraphics[height=6.3cm]{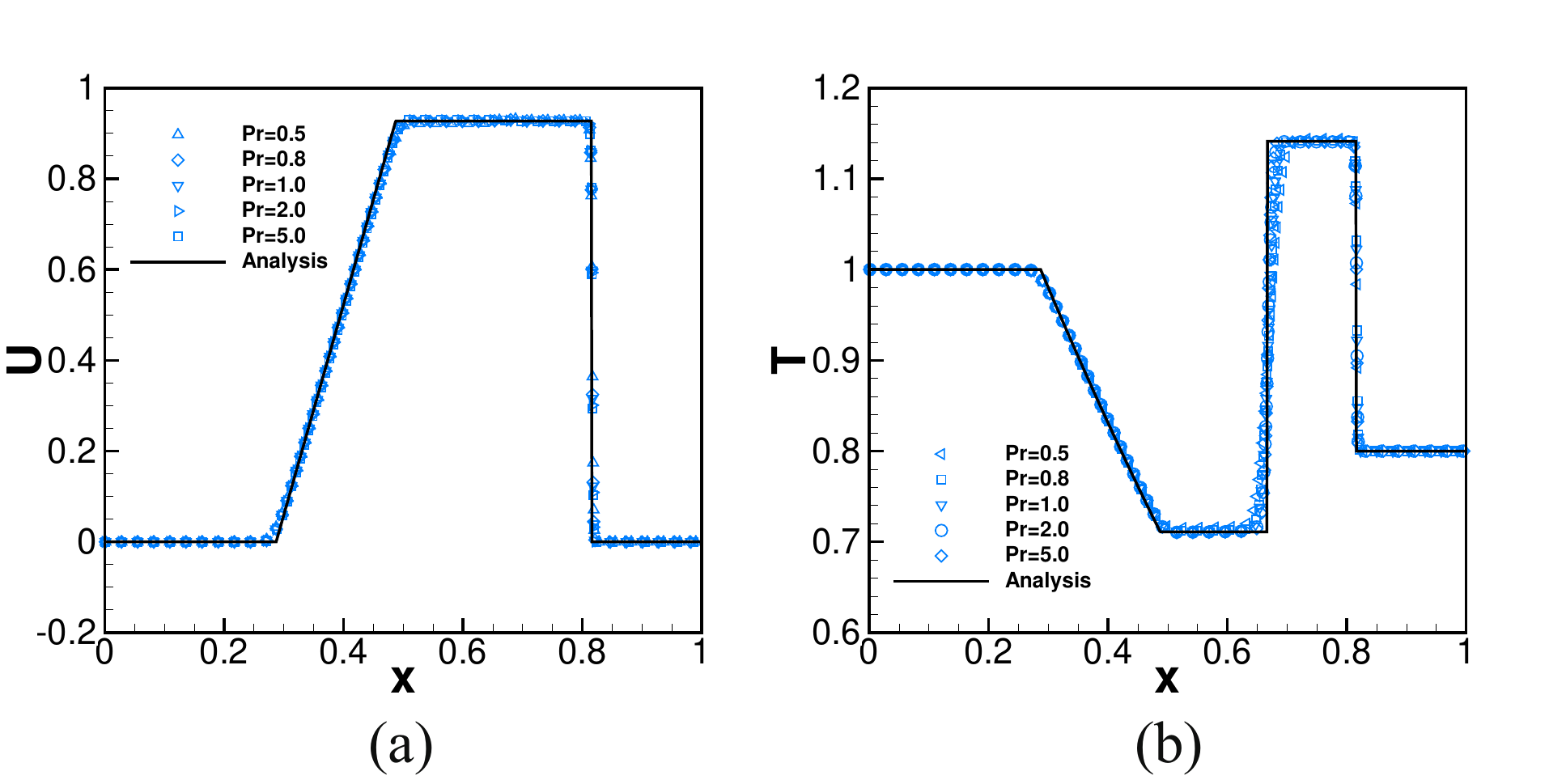}
\caption{Numerical simulation results by DBM and analytical solution of Sod shock tube. (a) velocity, (b) temperature in $\rm{Pr}=0.5,0.8,1.0,2.0$ and $5.0$. Symbols denote DBM simulation results in different $\rm{Pr}$ number and continuous black lines denote the corresponding analytical solutions. }
\label{Fig:sod1}
\end{figure*}

\begin{figure}[h]
\centering
\includegraphics[height=7cm]{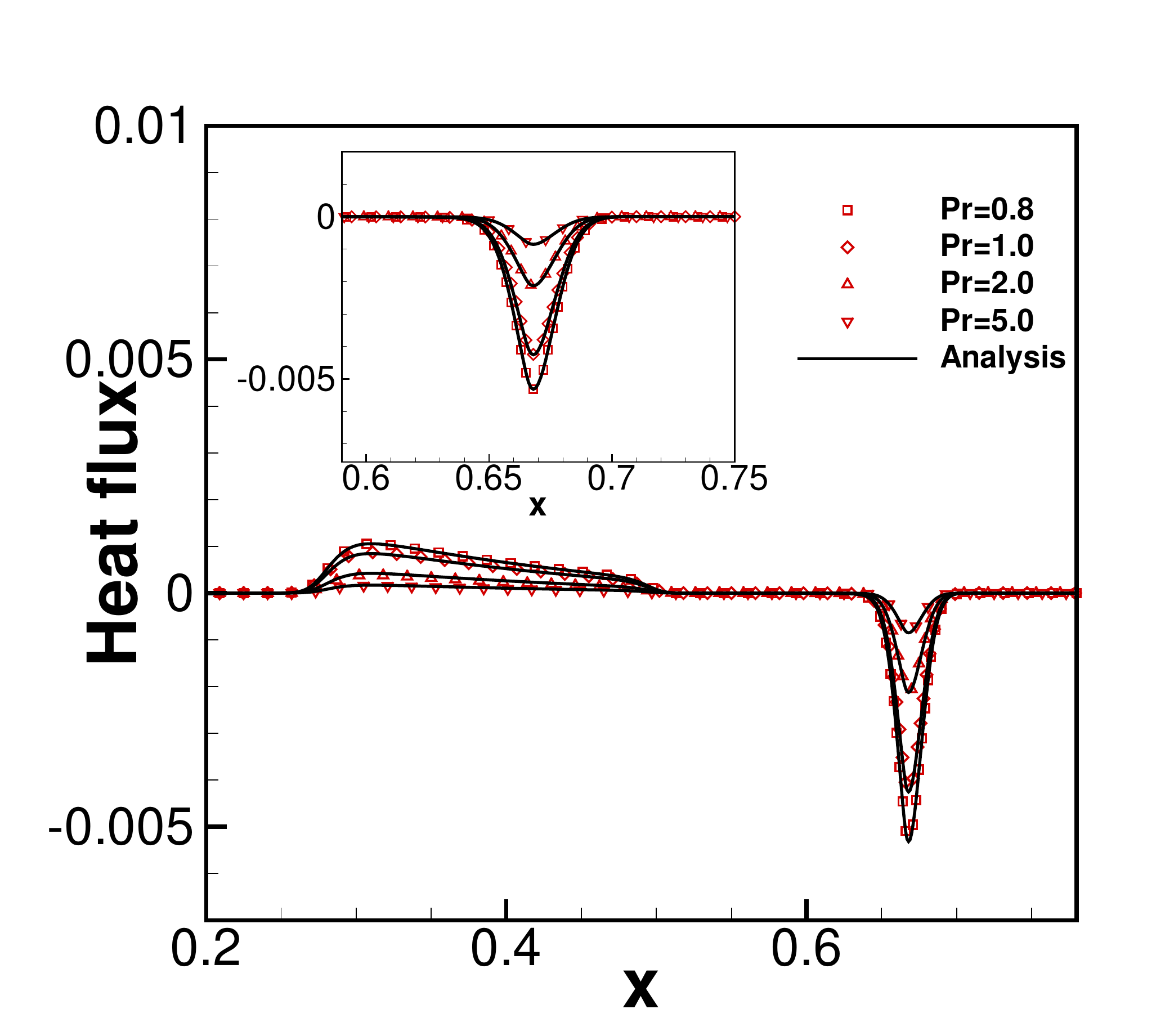}
\caption{TNE quantity  $ \Delta _{3,1x}^* $  and heat flux analytical solution in Pr=0.8,1.0,2.0 and 5.0. Symbols denote DBM simulation results in different Pr number and continuous lines denote the corresponding analytical solutions. }
\label{Fig:sod2}
\end{figure}

We select temperature-$x$ diagram at $t = 0.18$, and the results of grid-independence verification are shown in Fig.\ref{Fig:NX2}, and the grid mesh of simulation are $ Nx=200, 500, 1000$ and $2000$, respectively. The results have converged when the number of meshes $ Nx=1000$, so the grid mesh we choose are $ {Nx} \times {Ny} = 1000 \times 1 $ . Figure \ref{Fig:sod1} shows the comparison of the velocity and temperature between the DBM simulation results of the sod shock tube and the analytical solution in $ t = 0.18 $ , $\rm{Pr}$ number choosen in this case are 0.5, 0.8, 1.0, 2.0 and 5.0. The continuous lines denote the analytical solutions and the symbols denote DBM simulation results. Fig. \ref{Fig:sod1} shows  satisfying agreement between the two results. As an example, we calculate the L1-norms of physical quantities when $Pr=1$ and the temperature L1-norms is 4.69714, the velocity L1-norms is 3.99513. Furthermore, four different $\rm{Pr}$ numbers are choosen: $\rm{Pr}=0.8,1.0,2.0$ and $5.0$ in order to exhibit TNE quantity  $ \Delta _{3,1x}^* $ . The comparison of TNE quantity  $ \Delta _{3,1x}^* $ in the whole field with heat flux analytical solution in different $\rm{Pr}$ number is shown in Fig. \ref{Fig:sod2}. It shows a good fit in this figure. The above two numerical simulation cases prove the accuracy of the new model proposed.

\subsection{One-dimensional detonation wave propagation }

\begin{figure}[htbp]
\centering
\includegraphics[height=7cm]{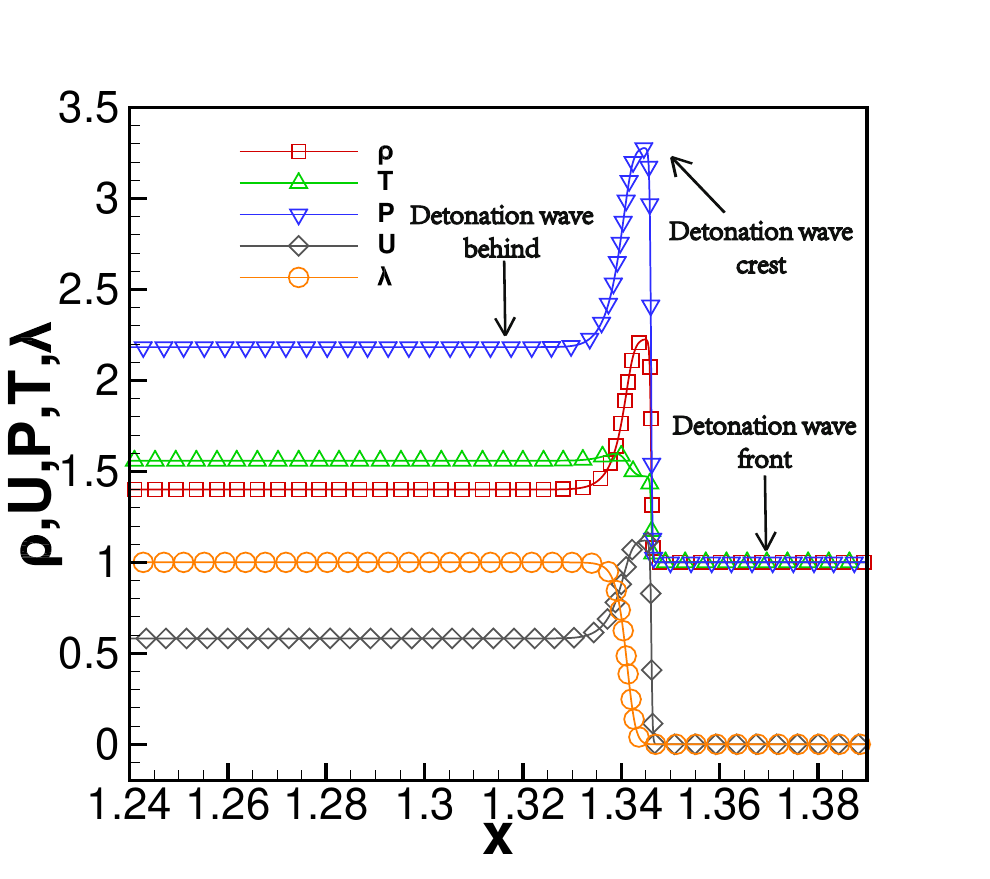}
\caption{Profiles of physical quantities  at $t=0.3$ with $\rm{Pr}= 1.0$, $\rm{Ma}  = 1.7444$. }
\label{Fig:odp1}
\end{figure}

\begin{figure}[htbp]
\centering
\includegraphics[height=4.0cm]{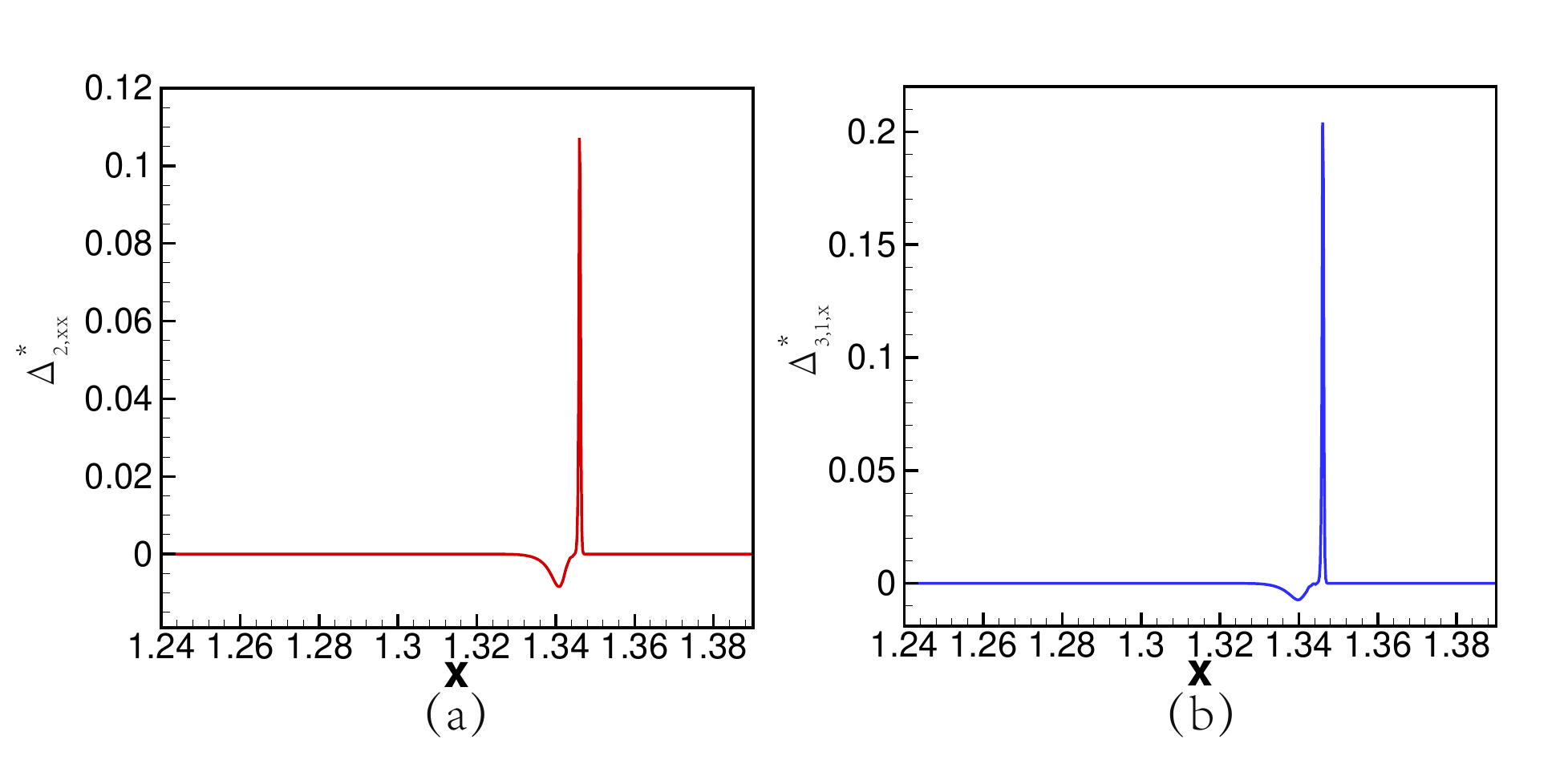}
\caption{Profiles of TNE quantities at $t=0.3$ with $\rm{Pr}= 1.0$, $\rm{Ma}  = 1.7444$. (a) $ \Delta _{2,xx}^* $, (b) $ \Delta _{3,1x}^* $ . }
\label{Fig:odp2}
\end{figure}

\begin{figure}[htbp]
\centering
\includegraphics[height=7cm]{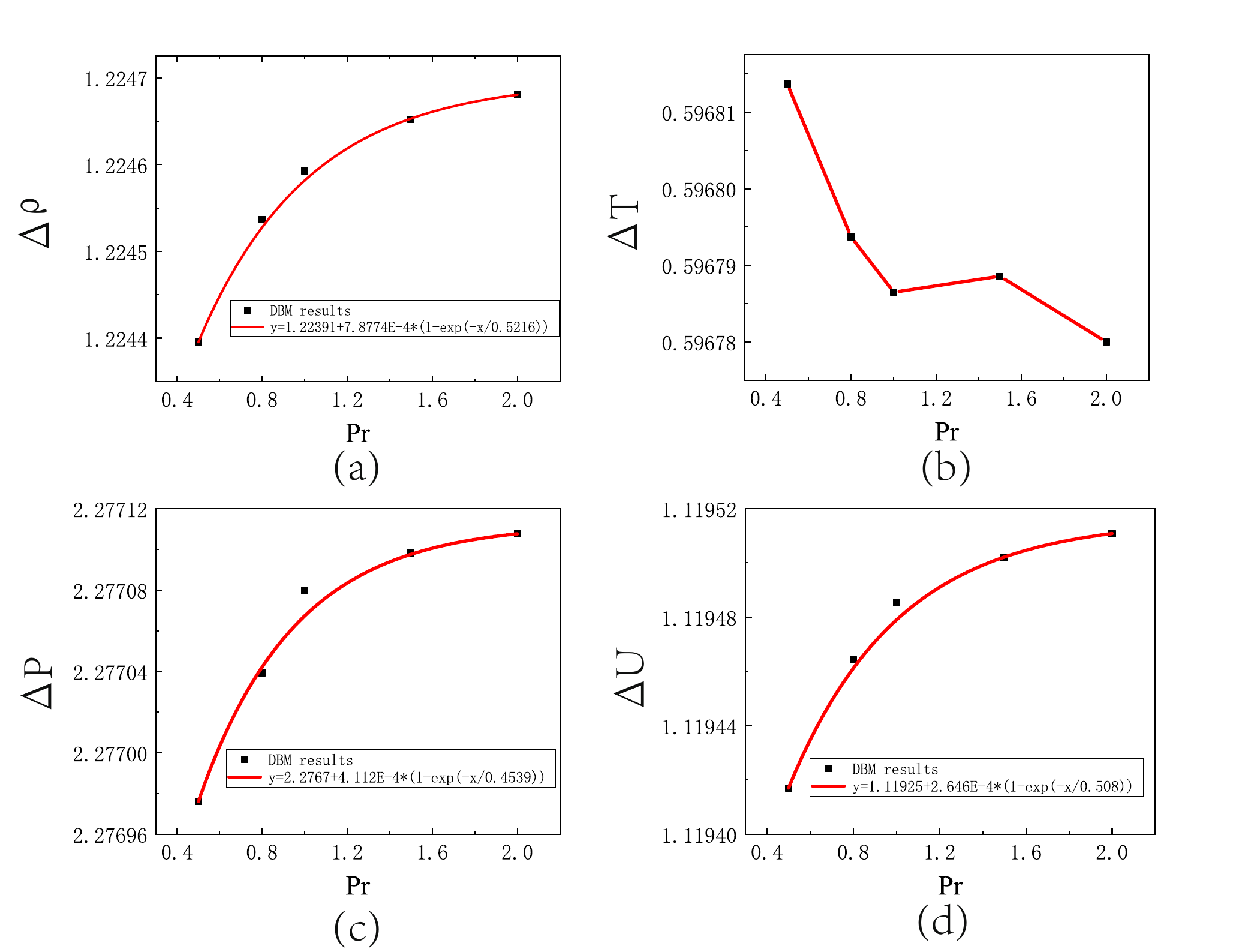}
\caption{Profiles of the difference between the detonation wave crest and wave front of physical quantities at $t=0.3$ with $\rm{Pr}= 0.5, 0.8,1.0,1.5 $ and $ 2.0$, $\rm{Ma}  = 1.7444$. (a) Density, (b) Temperature, (c) Pressure, (d) Velocity. }
\label{Fig:odp3}
\end{figure}

\begin{figure}[htbp]
\centering
\includegraphics[height=7cm]{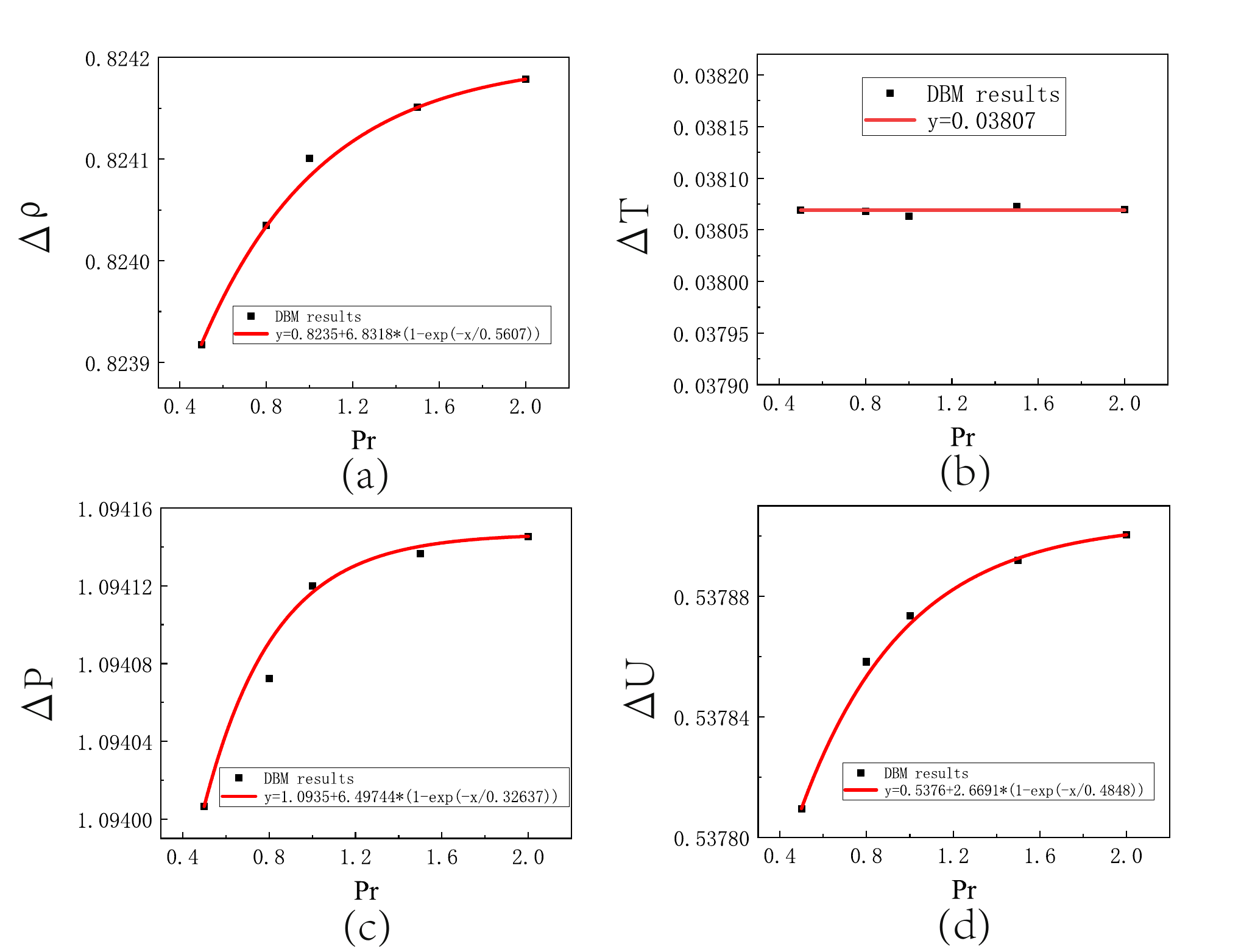}
\caption{Profiles of the difference between the detonation wave crest and wave behind of physical quantities at $t=0.3$ with $\rm{Pr}= 0.5, 0.8,1.0,1.5 $ and $ 2.0$, $\rm{Ma}  = 1.7444$. (a) Density, (b) Temperature, (c) Pressure, (d) Velocity. }
\label{Fig:odp7}
\end{figure}

\begin{figure}[htbp]
\centering
\includegraphics[height=3.6cm]{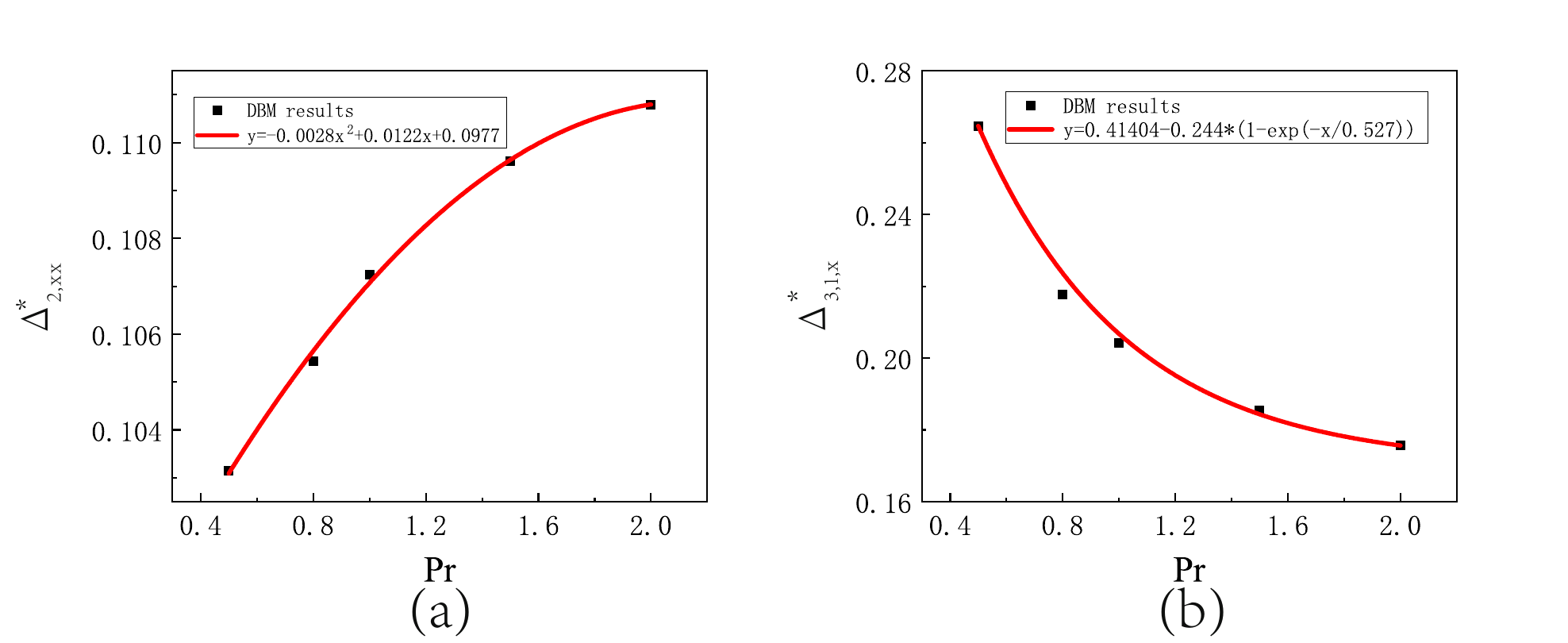}
\caption{Profiles of the difference between the detonation wave crest and wave front of TNE quantities at $t=0.3$ with $\rm{Pr}= 0.5, 0.8, 1.0, 1.5 $ and $ 2.0$, $\rm{Ma}  = 1.7444$. (a) $ \Delta _{2,xx}^* $, (b) $\Delta _{3,1x}^*$. }
\label{Fig:odp5}
\end{figure}

\begin{figure}[htbp]
\centering
\includegraphics[height=3.6cm]{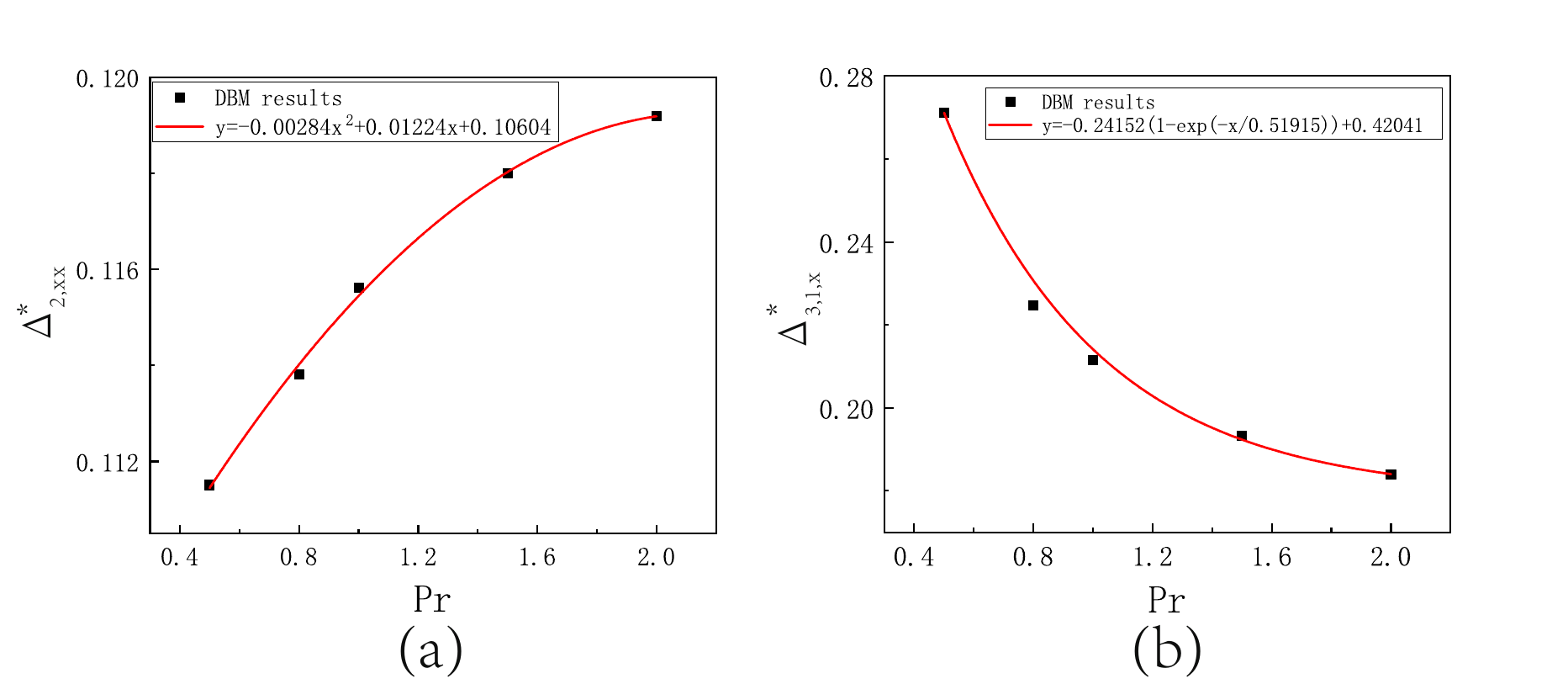}
\caption{Profiles of the difference between the detonation wave crest and wave behind of TNE quantities at $t=0.3$ with $\rm{Pr}= 0.5, 0.8, 1.0, 1.5 $ and $ 2.0$, $\rm{Ma}  = 1.7444$. (a) $ \Delta _{2,xx}^* $, (b) $\Delta _{3,1x}^*$. }
\label{Fig:odp9}
\end{figure}

\begin{figure}[htbp]
\centering
\includegraphics[height=7cm]{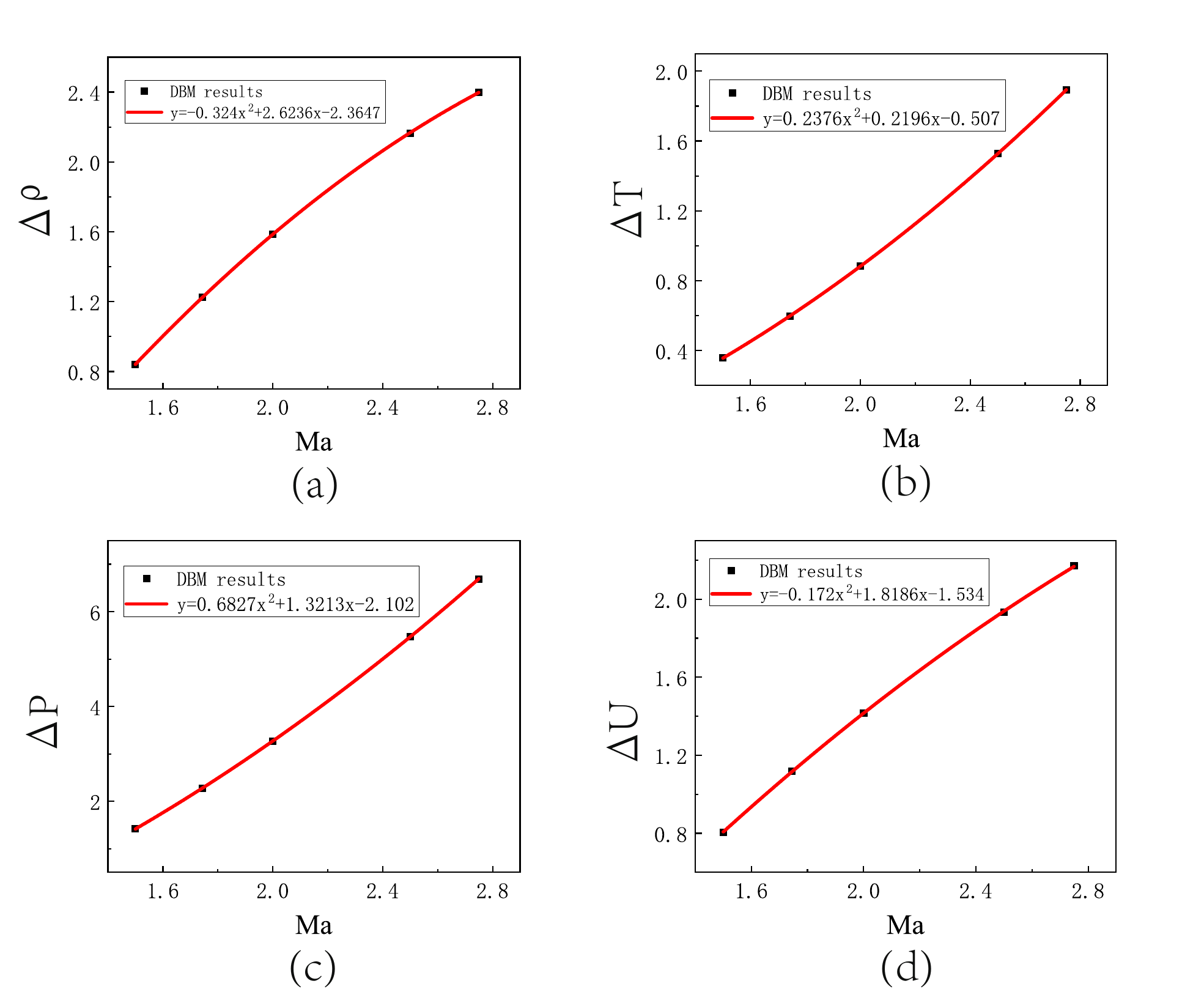}
\caption{Profiles of the difference between the detonation wave crest and wave front of physical quantities with $\rm{Ma}= 1.5, 1.7444, 2.0, 2.5 $ and $ 2.75$, $\rm{Pr}  = 1.0$. (a) Density, (b) Temperature, (c) Pressure, (d) Velocity. }
\label{Fig:odp4}
\end{figure}

\begin{figure}[htbp]
\centering
\includegraphics[height=7cm]{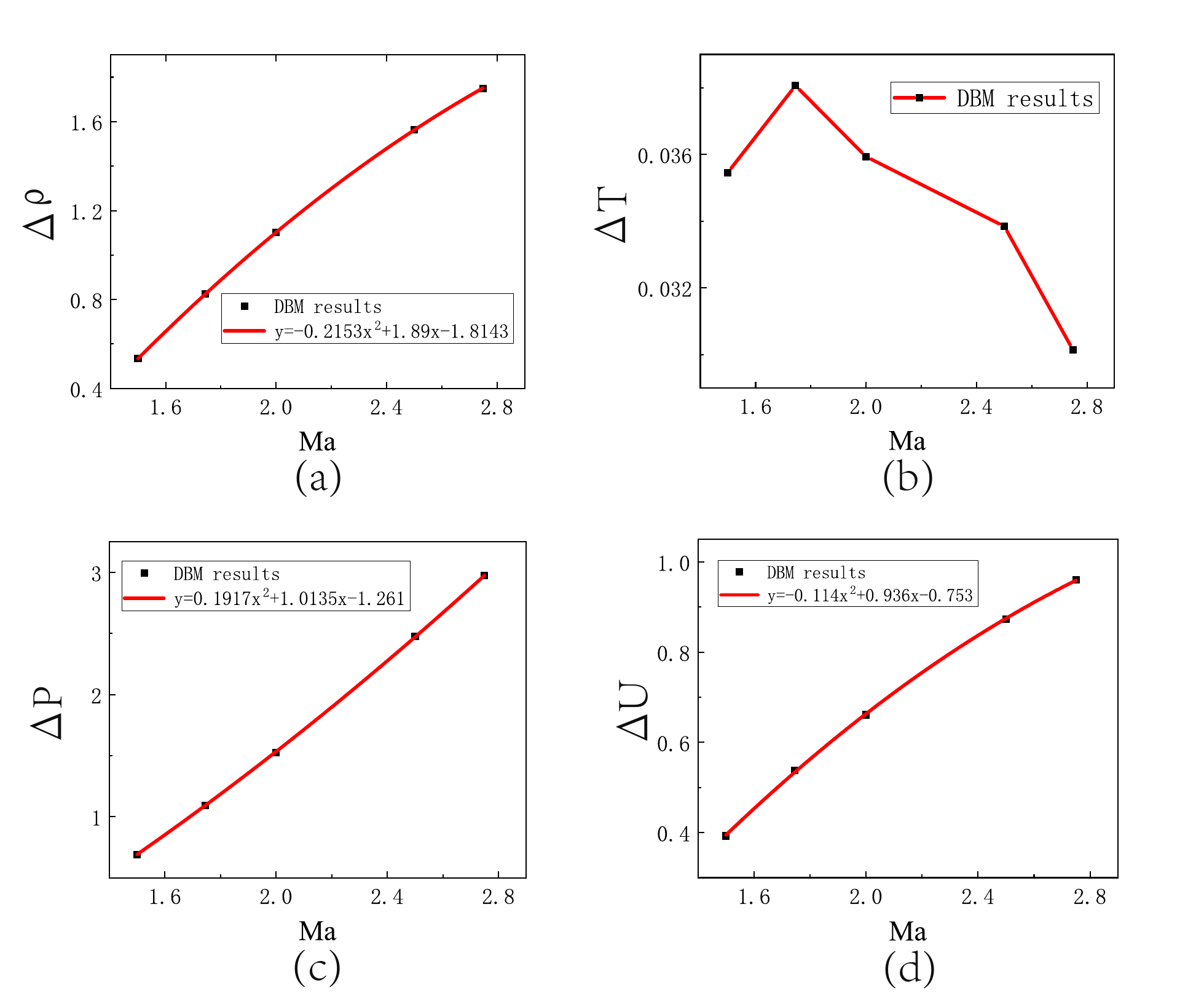}
\caption{Profiles of the difference between the detonation wave crest and wave behind of physical quantities with $\rm{Ma}= 1.5, 1.7444, 2.0, 2.5 $ and $ 2.75$, $\rm{Pr}  = 1.0$. (a) Density, (b) Temperature, (c) Pressure, (d) Velocity. }
\label{Fig:odp8}
\end{figure}

\begin{figure}[htbp]
\centering
\includegraphics[height=3.9cm]{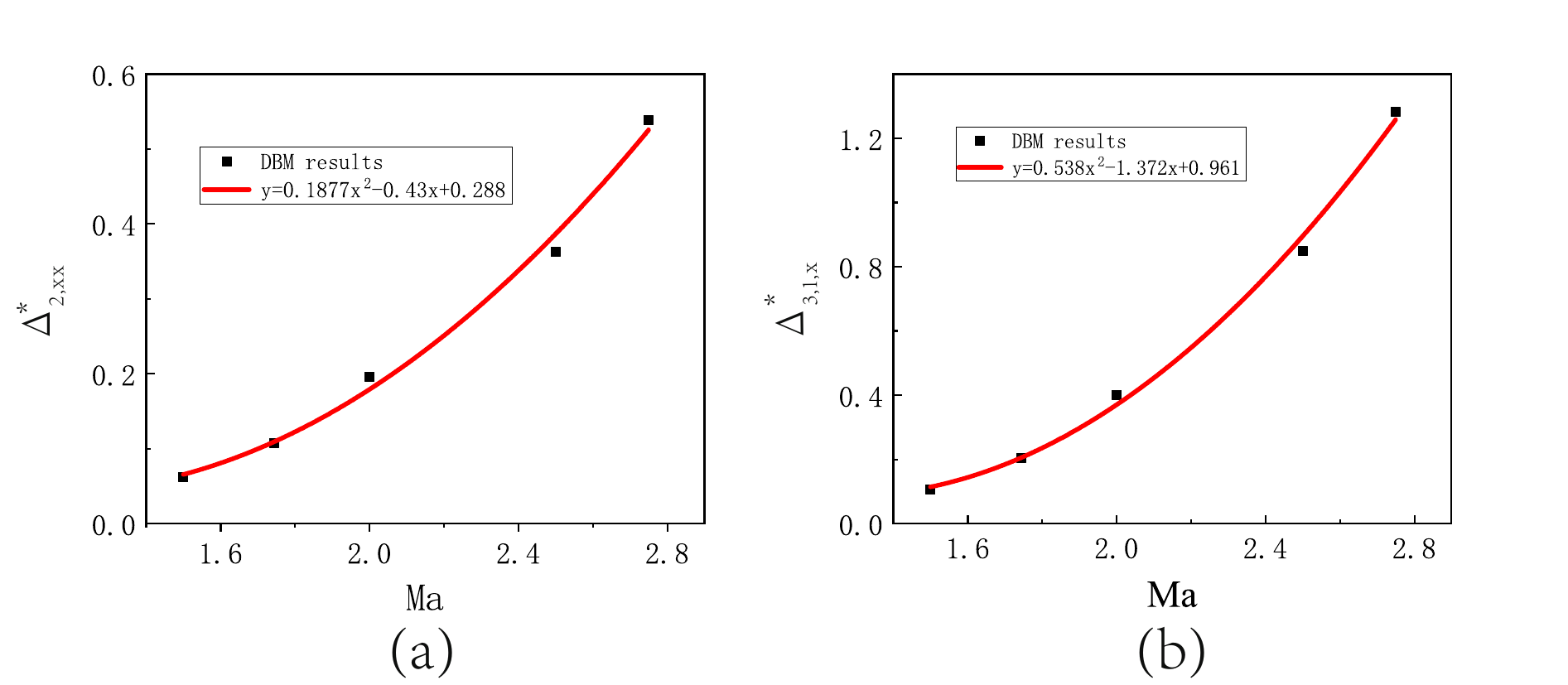}
\caption{Profiles of the difference between the detonation wave crest and wave front of TNE quantities with $\rm{Ma}= 1.5, 1.7444, 2.0, 2.5 $ and $ 2.75$, $\rm{Pr}  = 1.0$. (a) $ \Delta _{2,xx}^* $, (b) $\Delta _{3,1x}^*$. }
\label{Fig:odp6}
\end{figure}

\begin{figure}[htbp]
\centering
\includegraphics[height=3.9cm]{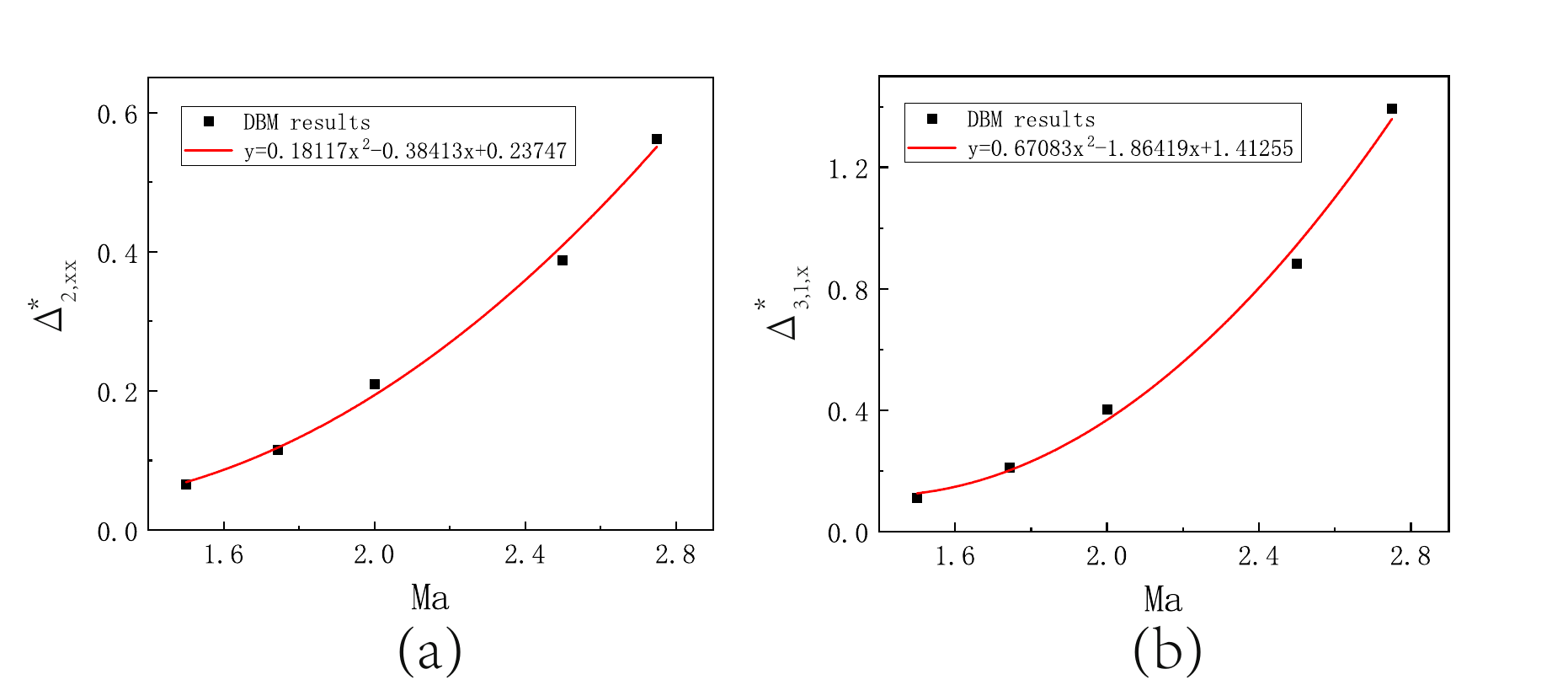}
\caption{Profiles of the difference between the detonation wave crest and wave behind of TNE quantities with $\rm{Ma}= 1.5, 1.7444, 2.0, 2.5 $ and $ 2.75$, $\rm{Pr}  = 1.0$. (a) $ \Delta _{2,xx}^* $, (b) $\Delta _{3,1x}^*$. }
\label{Fig:odp10}
\end{figure}

Simulation of three cases which are including chemical reaction using above DBM model will be carried out in the following. One-dimensional detonation wave propagation along the positive $x$-axis will be carried out. Initially, the physical quantities of the flow field are
\begin{equation}\label{Eq:RES-10}
\begin{aligned}
{(\rho ,u,v,T,\lambda )_L} &= (1.38837,0.57735,0,1.57856,1), \\
{(\rho ,u,v,T,\lambda )_R} &= (1,0,0,1,0).
\end{aligned}
\end{equation}
Where the subscript $ L $ and $ R $ represent left and right parts of the flow field, respectively. The grid mesh of flow field is ${Nx} \times {Ny} = 7500 \times 1$ , the spatial step and temporal step are $ \Delta x = \Delta y = 0.0002 $  and  $ \Delta t = 5 \times {10^{ - 6}} $. $ \tau  = 1 \times {10^{ - 5}} $, $ Q = 1.0 $, reaction rate coefficient $k = 1000$ , Mach number $\rm{Ma}  = 1.7444$, and blasting temperature is $ {T_{th}} = 1.1 $, $ \rm{Pr}  = 1.0 $, $ \gamma  = 7/5 $. As for boundary condition, inflow and outflow boundary conditions are adopted. The profiles of density, temperature, velocity, pressure, and $\lambda $ of the physical system at $t$ = 0.3 with $ \rm{Pr}  = 1.0 $ are shown in Fig. \ref{Fig:odp1}. Figure \ref{Fig:odp2} is the TNE quantities $ \Delta _{2,xx}^* $ and  $ \Delta _{3,1x}^* $ at $t=0.3$ with $\rm{Pr}= 1.0$. Meanwhile, we calculate the difference between the detonation wave crest and wave front and the difference between the detonation wave crest and wave behind of physical quantities (density, temperature, pressure and velocity) at $t=0.3$ with $\rm{Pr}= 0.5, 0.8,1.0,1.5 $ and $ 2.0$ , which are shown in Fig. \ref{Fig:odp3} and Fig. \ref{Fig:odp7}, respectively. And similarly, the difference of TNE quantities are shown in Fig. \ref{Fig:odp5} and Fig. \ref{Fig:odp9}. It can be seen that each physical quantities change monotonously with the increase of Prandtl numbers. And as the Prandtl number increases, the viscous stress $ \Delta _{2,xx}^* $ increases slightly, but the heat flow $ \Delta _{3,1x}^* $ decreases significantly. Meanwhile, the influence of Mach numbers on the evolution of one-dimensional detonation wave propagation is also investigated. We conduct five cases with various Mach numbers, $\rm{Ma}=1.5, 1.7444, 2.0, 2.5$ and $2.75$, respectively. Figure \ref{Fig:odp4} shows the difference between the detonation wave crest and wave front of physical quantities with different $\rm{Ma}$ numbers and figure \ref{Fig:odp8} shows the difference between the detonation wave crest and wave behind. In addition, figures \ref{Fig:odp6} and \ref{Fig:odp10} shows the difference of TNE quantities. It can be observed that, the Pr number, Mach number, detonation heat and intensity coefficient of reaction rate all affect the width, height and even morphology of von Neumann peak. The details are as follows:

(i) (within the range of numerical experiments) the peak heights of pressure, density and flow velocity increase exponentially with the number of Pr.
The maximum stress increases parabolically with Pr number, and the maximum heat flux decreases exponentially with Pr number.
(ii) the peak heights of pressure, density, temperature and flow velocity and the maximum stress within the peak are parabolically increase with the Mach number, the maximum heat flux decreases exponentially with the Mach number.


\subsection{The collision of detonation wave and shock wave}

\begin{figure}[htbp]
\centering
\includegraphics[height=7cm]{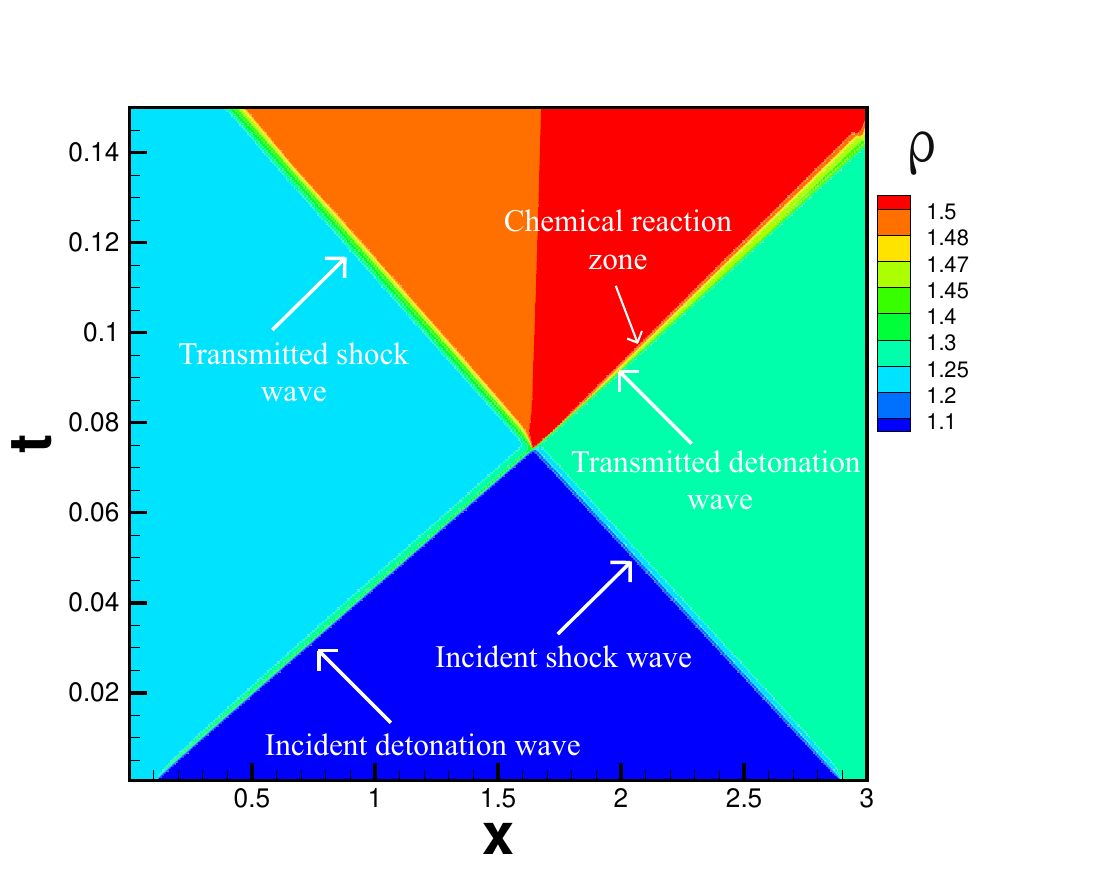}
\caption{Density contour in the location coordinate $x$-$t$ coordinate system.}
\label{Fig:cds1}
\end{figure}

\begin{figure}[htbp]
\centering
\includegraphics[height=7cm]{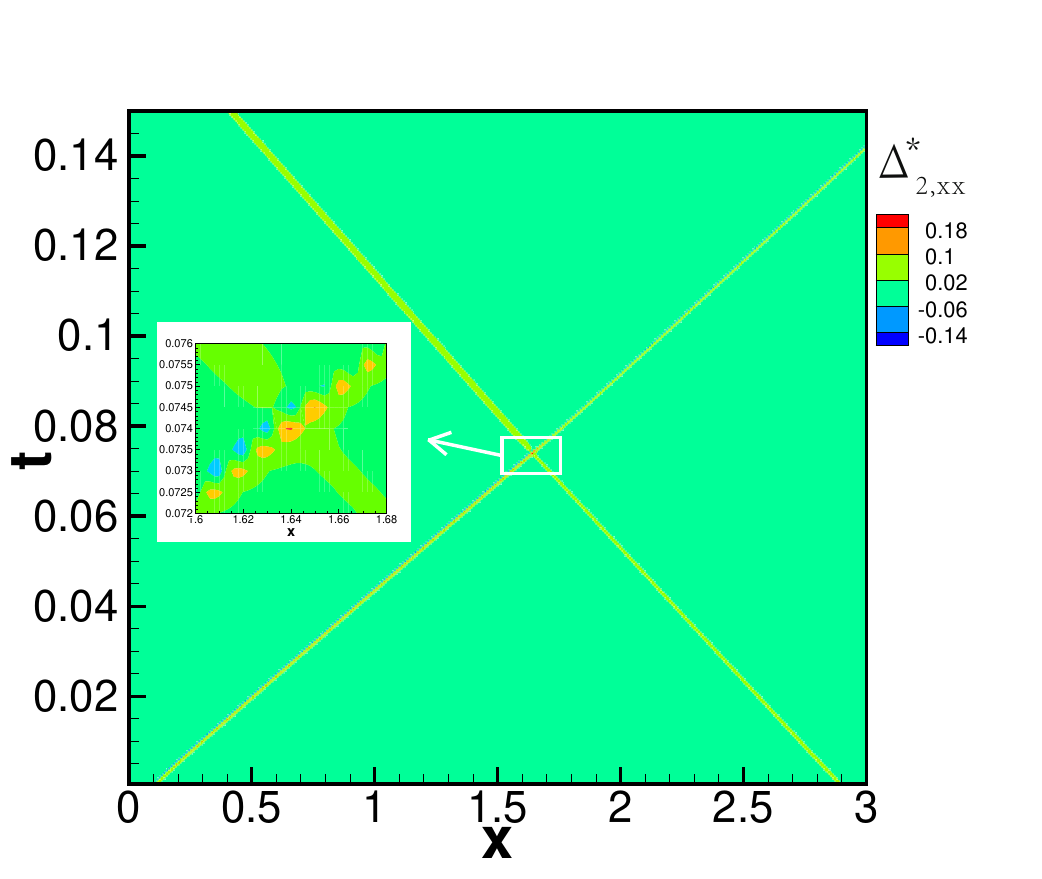}
\caption{TNE quantity $\Delta _{2,xx}^*$ contours in the location coordinate $x$-$t$ coordinate system.}
\label{Fig:cds2}
\end{figure}

\begin{figure}[htbp]
\centering
\includegraphics[height=7cm]{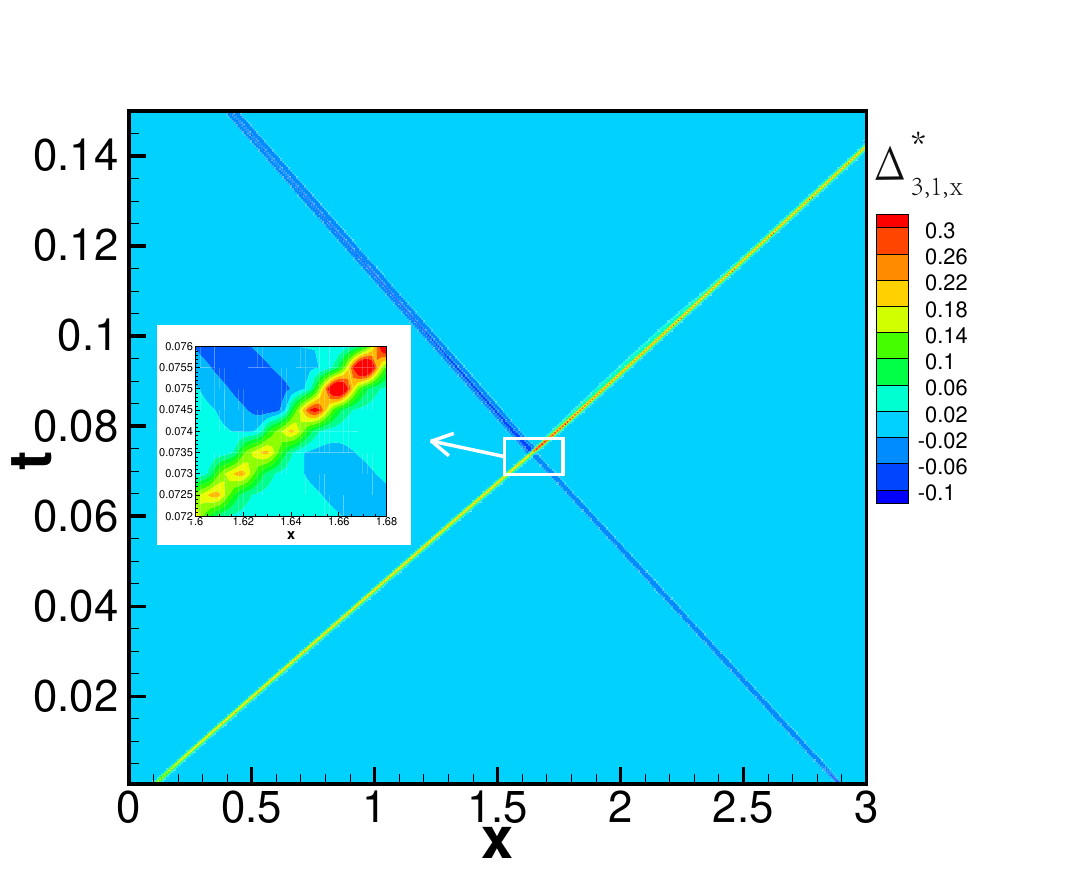}
\caption{TNE quantity $\Delta _{3,1x}^*$ contours in the location coordinate $x$-$t$ coordinate system.}
\label{Fig:cds3}
\end{figure}

The interaction of detonation waves and shock waves occurs widely in the industrial field, and research on this has practical significance. A one-dimensional stable detonation wave and a shock wave propagate reversely along the $x$-axis, and they are to impact at somewhere. For the computational domain, the physical quantity of the flow field at the initial moment are
\begin{equation}\label{Eq:RES-6}
{(\rho ,{u},{v},T,\lambda )_L} = (1.22727,0.39284,0,1.49383,1) ,
\end{equation}
\begin{equation}\label{Eq:RES-7}
{(\rho ,{u},{v},T,\lambda )_M} = (1,0,0,1,0) ,
\end{equation}
\begin{equation}\label{Eq:RES-8}
{(\rho ,{u},{v},T,\lambda )_R} = (1.25581,0.34570,0,1.26346,0) ,
\end{equation}
where the subscript $ L $ ,  $ M $ and $ R $ represent the left, middle and right part of the flow field, respectively. The Mach number of detonation wave is $\rm{Ma} = 1.5 $ ,and the Mach number of shock wave is $\rm{Ma} = 1.2 $ , $ Q = 0.2315 $ , $ Nx \times Ny = 1500 \times 1 $, and the blasting temperature is $ {T_{th}} = 1.3 $, $k = 1000$ .The spatial step is $ \Delta x = \Delta y = 0.002 $ , and the temporal step is $ \Delta t = 5 \times {10^{ - 5}} $ , $ \tau  = 0.002 $ , $ \Pr  = 1 $ . As for boundary condition, the periodic boundary condition is adopted in the $y$ direction, and the outflow boundary condition is adopted in the $x$ direction. Figure \ref{Fig:cds1} exhibits the density of the whole flow field in the location coordinate $x$ - time $t$ diagram. The incident detonation wave, incident shock wave, transmitted detonation wave, transmitted shock wave and chemical reaction zone are captured by DBM clearly. To investigate the TNE behaviors, Figs. \ref{Fig:cds2} and \ref{Fig:cds3} exhibit TNE quantities  $ \Delta _{2,xx}^* $ and $\Delta _{3,1x}^*$ of this case, It is clear that TNE information  $ \Delta _{2,xx}^* $ and $\Delta _{3,1x}^*$ have larger values on the wave front of detonation waves and shock waves than else in the flow field, and the reason for this is that the gradient of macroscopic physical quantities is larger, and the degree of non-equilibrium is higher. It can be observed that the value of $ \Delta _{2,xx}^* $ is maximum when detonation wave and shock wave collide in Fig. \ref{Fig:cds2}, and the similar situation can also be found in Fig. \ref{Fig:cds3}. And at the detonation wave, the value of $ \Delta _{2,xx}^* $  become less after colliding, but we can see the values of $ \Delta _{2,xx}^* $ for shock wave become greater by comparing the value of $ \Delta _{2,xx}^* $ between after and before collision. Moreover, as shown in Fig. \ref{Fig:cds3}, the values of $\Delta _{3,1x}^*$ at detonation wave and shock wave become greater than before. In other words, the collision of detonation wave and shock wave causes an increase in the heat flow at the detonation wave and shock wave in this case.

\subsection{Two-dimensional detonation problem}
To investigate the influence of $\rm{Pr}$ number on detonation, numerical simulation of a two-dimensional detonation problem is performed. In a square computational domain with $ Nx \times Ny = 400 \times 400 $ , two detonation waves occur in the upper left and lower left plot $A$ and $B$ of the computational domain, respectively. Detonation waves will spread and collide at a certain moment. Initial conditions of flow field are as follows,
\begin{equation}\label{Eq:RES-9}
\left( {\rho ,u,v,T,\lambda } \right) = \left\{ {\begin{array}{*{20}{c}}
{\left( {1.388,0.577,0,1.579,1} \right)}&{{\rm{ A,B}}}\\
{\left( {1,0,0,1,0} \right)}&{{\rm{others}}}
\end{array}} \right.
\end{equation}

\begin{figure*}[htbp]
\centering
\includegraphics[height=7cm]{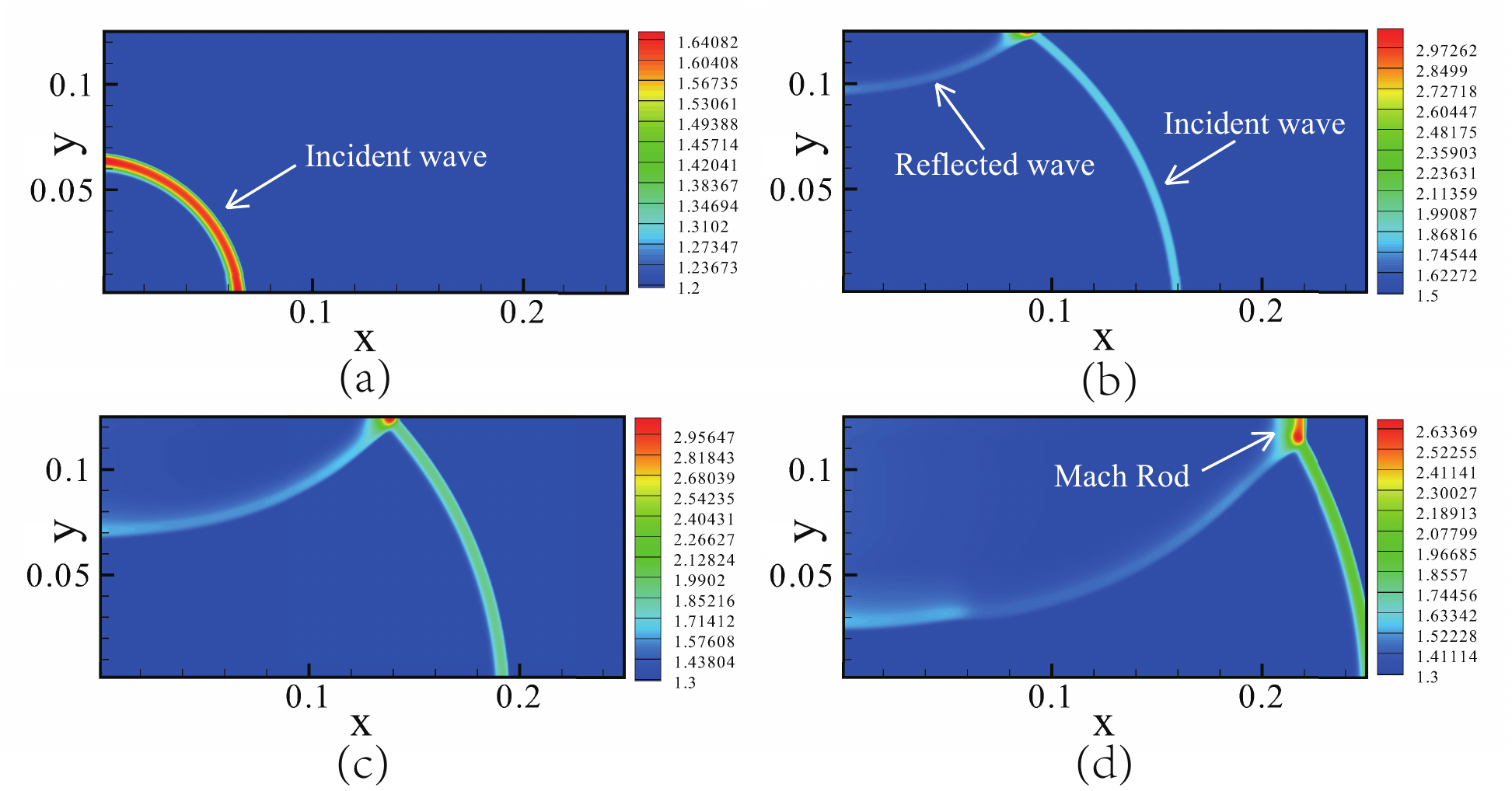}
\caption{Profiles of density with $\rm{Pr}=1.0$. (a) $t=0.045$, (b) $t=0.1$, (c)$t=0.118$, (d) $t=0.15$.}
\label{Fig:tdp1}
\end{figure*}
\begin{figure*}[htbp]
\centering
\includegraphics[height=7cm]{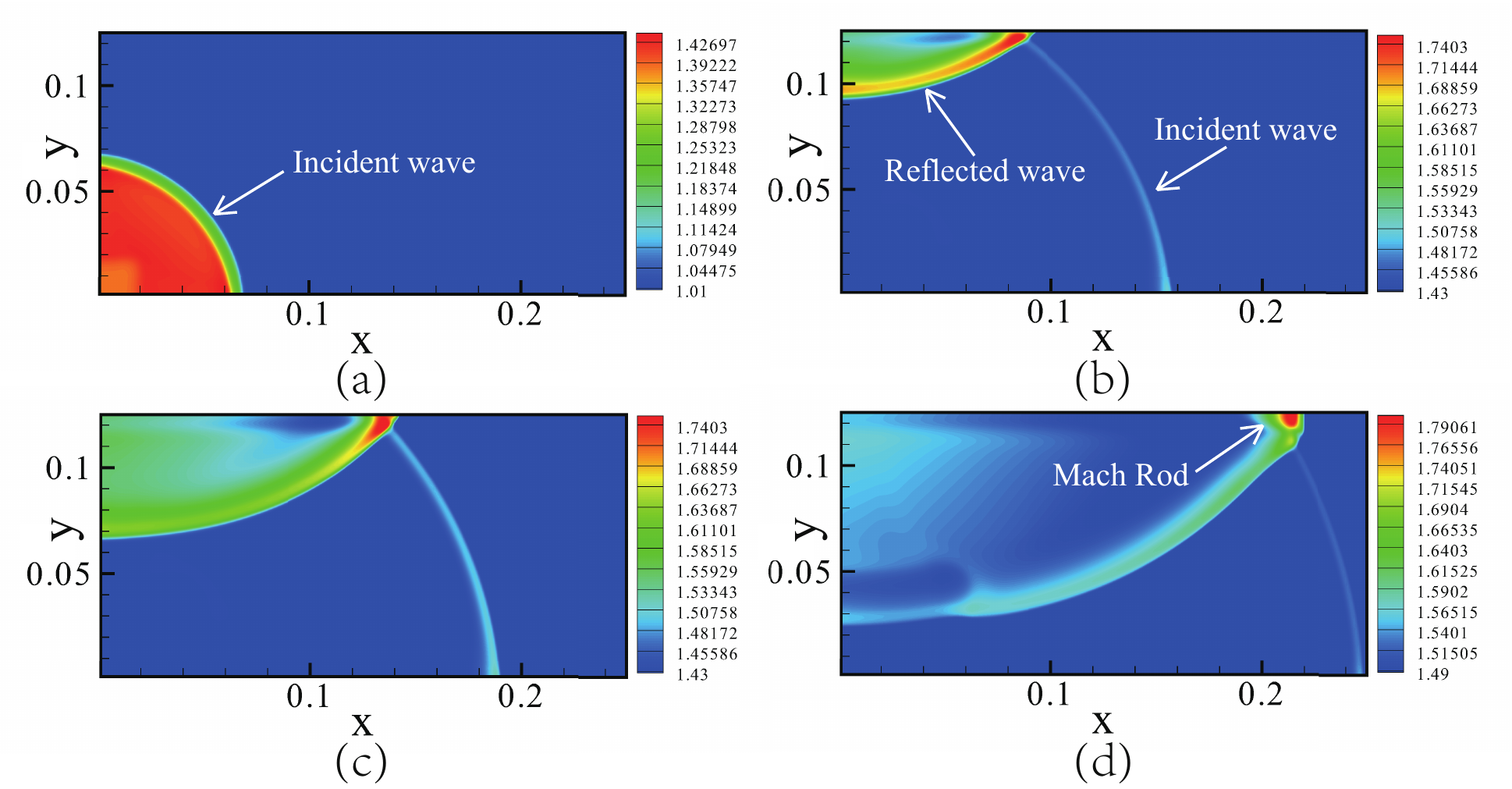}
\caption{Profiles of temperature with $\rm{Pr}=1.0$. (a)$t=0.045$, (b) $t=0.1$, (c) $t=0.118$, (d)$t=0.15$.}
\label{Fig:tdp2}
\end{figure*}
\begin{figure*}[htbp]
\centering
\includegraphics[height=7cm]{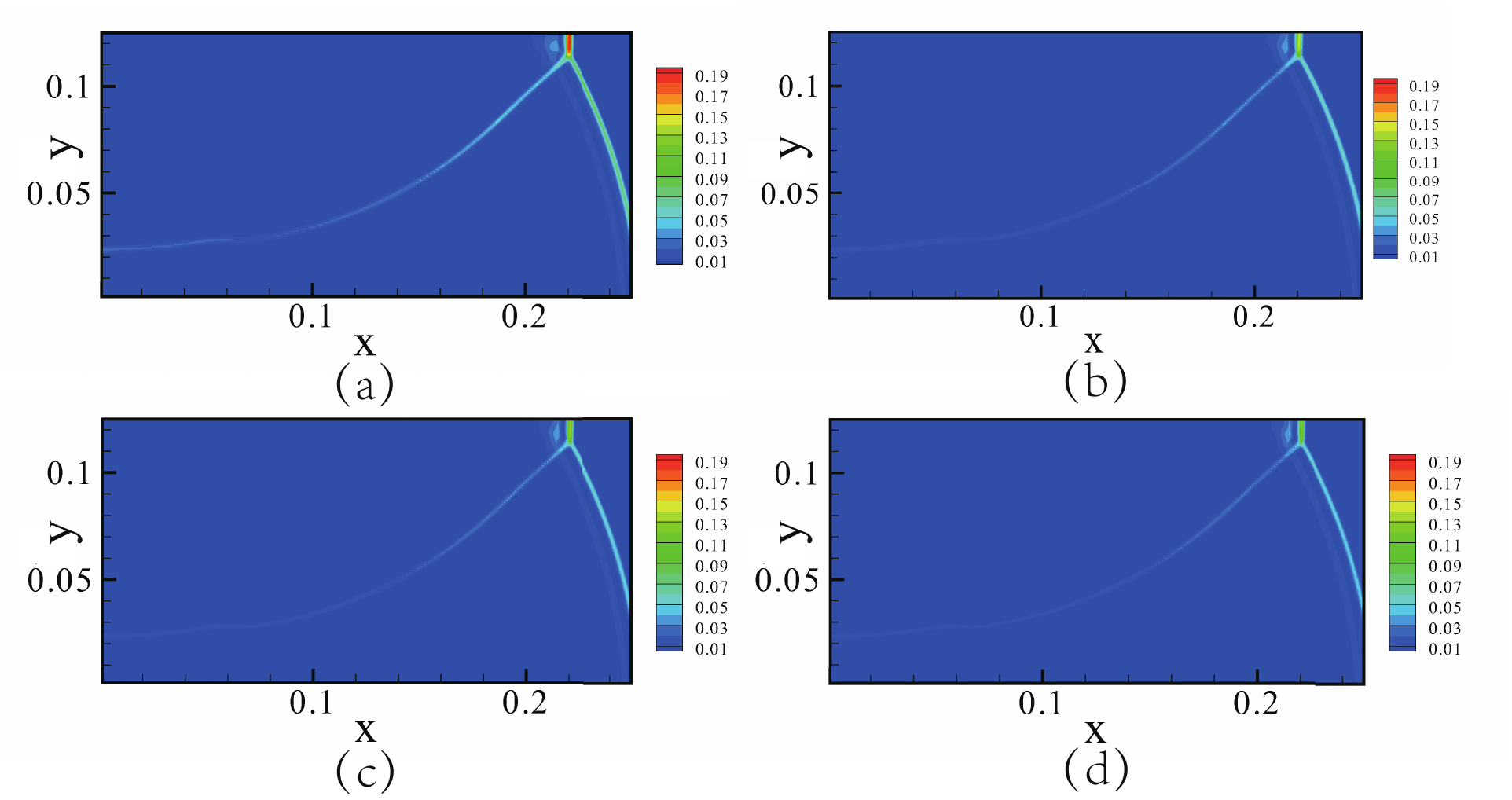}
\caption{Profiles of $ \left| {\Delta _{3,1}^{\rm{*}}} \right| $ at $t=0.15$. (a) $\rm{Pr}= 0.5$, (b) $\rm{Pr}= 0.8$, (c) $\rm{Pr}= 1.0$, (d) $\rm{Pr}= 1.5$.}
\label{Fig:tdp3}
\end{figure*}


Parameters are as follows:  $ \Delta x = \Delta y = 0.001 $ , $ \Delta t = 1 \times {10^{ - 5}} $ , $ \tau {\rm{ = }}1 \times {10^{-5}} $ , $ Q = 1.0 $ , $k = 1000$, specific heat ratio  $ \gamma  = 1.4 $ , $ {T_{th}} = 1.1 $ . In this case, free outflow boundary conditions are adopted on the four sides of computational domain, respectively. Due to the symmetry of the flow field, only the lower half of the simulation area is shown in the figures. From Figs. \ref{Fig:tdp1} and \ref{Fig:tdp2}, it can be seen that the density and temperature contours of physical system on the evolution at four different times, with $\rm{Pr} =1$. Detonation waves spread and collide in the flow field, and we can find the Mach rod in Fig. \ref{Fig:tdp1} (d). The reflected wave angles of Fig. \ref{Fig:tdp1} (b) $25.1^\circ $ agree well with analytic angle $22.3^\circ $.  Meanwhile, TNE quantities in $\rm{Pr}=0.5$, $0.8$, $1.0 $ and $1.5$ at $ t = 0.15 $ are compared. The values of  $ \Delta _{2,xx}^* $ change slightly in various $\rm{Pr}$ numbers on the whole, and the TNE quantities $ \Delta _{2,xy}^* $ and $ \Delta _{2,yy}^* $ are analogous. Shown in Fig. \ref{Fig:tdp3} is the intensity of the heat flux $ \left| {\Delta _{3,1}^{\rm{*}}} \right| $ contours of detonation waves, and it can be seen that TNE quantities $\left| {\Delta _{3,1}^{\rm{*}}} \right| $ of detonation waves increases with the decrease of $\rm{Pr}$ number. Besides, from Fig. \ref{Fig:tdp3}, it is shown that the speed of detonation waves in different $\rm{Pr}$ numbers is unchanged basically. This indicates that the propagation speed of the detonation waves is not affected evidently by the change of Pr number.

\section{Conclusion and discussion}

A discrete Boltzmann model including chemical reaction based on Shakhov model is proposed. The contribution of chemical reaction enters the model via modifying the collision term, more specifically, modifying the local temperature which determines the local equilibrium distribution function ${f^{eq}}$. Compared with the Navier-Stokes model, the proposed model can better capture the fine physical structure of detonation wave. Compared with the previous model based on BGK model, the current DBM based on Shakhov model possesses flexible Prandtl number, which significantly extends the application range. It should be noticed that the chemical reaction rate model choosen in this paper is just an example, and people can choose the chemical reaction rate model according to practical need. The model is validated by well-known benchmarks. Non-equilibrium behaviors of detonations with various Prandtl numbers and Mach numbers are numerically investigated.
The von Neumann peak relative to the front platform and the von Neumann peak relative to the back platform, just as the kinetic moment about particle velocity $\bf{v}$ and the central kinetic moment which is about $(\bf{v}-\bf{u})$,  are descriptions of the same system and the same behavior from different perspectives, where $\bf{u}$ is mean velocity.
The central kinetic moment, describes the kinetic characteristics caused by pure thermal motion (fluctuation), while the kinetic moment about $\bf{v}$ contains the contribution of the mean velocity $\bf{u}$.
In the case of detonation wave, the von Neumann peak is a behavior that cannot be given by the Hugoniot relations.  The existence of this peak is a typical non-equilibrium characteristic, and the region covered by the peak is non-equilibrium region.
The Prandtl number, Mach number, detonation heat and intensity coefficient of reaction rate all affect the width, height and even morphology of von Neumann peak.
The von Neumann peak relative to the wavefront platform contains some of the properties, such as the differences of density, pressure, temperature and flow velocity at the wave back and front, that the Hugoniot relations can describe.
This paper conducts a preliminary study on the von Neumann peak relative to the wave front and finds that
(i) (within the range of numerical experiments) the peak heights of pressure, density and flow velocity increase exponentially with the Prandtl number.
The maximum stress increases parabolically with Prandtl number, and the maximum heat flux decreases exponentially with Prandtl number.
(ii) The peak heights of pressure, density, temperature and flow velocity and the maximum stress within the peak are parabolically increase with Mach number.
The  caharacteristics of von Neumann peak relative to wave back can be studied in a similar way.

\begin{acks}

The authors thank Yanbiao Gan, Chuandong Lin, Ge Zhang, Jiahui Song, and Dejia Zhang, on helpful discussions.
This work was supported by the National Natural Science Foundation of China (under Grant Nos. 11772064, 12102397 and 11975053),
the opening project of State Key
Laboratory of Explosion Science and Technology (Beijing Institute of Technology) (under Grant No. KFJJ21-16M),
the China Postdoctoral Science Foundation (under Grant No. 2019M662521), the Natural Science Foundation of Shandong Province (under Grant No. ZR2020MA061),
and Shandong Province Higher Educational Youth Innovation Science and Technology Program (under Grant No. 2019KJJ009).

\end{acks}


\begin{thebibliography}{10}
\providecommand{\url}[1]{\texttt{#1}}
\providecommand{\urlprefix}{URL }
\expandafter\ifx\csname urlstyle\endcsname\relax
  \providecommand{\doi}[1]{DOI:\discretionary{}{}{}#1}\else
  \providecommand{\doi}{DOI:\discretionary{}{}{}\begingroup
  \urlstyle{rm}\Url}\fi
\providecommand{\eprint}[2][]{\url{#2}}

\bibitem{ren2014numerical}
Ren Z, Lu Z, Hou L et~al.
\newblock Numerical simulation of turbulent combustion: Scientific challenges.
\newblock \emph{Science China Physics, Mechanics \& Astronomy} 2014; 57(8):
  1495--1503.

\bibitem{ren2019supersonic}
Ren Z, Wang B, Xiang G et~al.
\newblock Supersonic spray combustion subject to scramjets: Progress and
  challenges.
\newblock \emph{Progress in Aerospace Sciences} 2019; 105: 40--59.

\bibitem{Ju-Review}
Ju Y.
\newblock Recent progress and challenges in fundamental combustion research.
\newblock \emph{Advances in Mechanics} 2014; 44(20): 201402.

\bibitem{Yang2013Large}
Yang Y, Wang H, Pope SB et~al.
\newblock Large-eddy simulation/pdf modeling of a non-premixed co/h2 temporally
  evolving jet flame.
\newblock \emph{Proceedings of the Combustion Institute} 2013; 34(1):
  1241--1249.

\bibitem{wang2020laminar}
Wang Y, Movaghar A, Wang Z et~al.
\newblock Laminar flame speeds of methane/air mixtures at engine conditions:
  performance of different kinetic models and power-law correlations.
\newblock \emph{Combustion and Flame} 2020; 218: 101--108.

\bibitem{wang2021effects}
Wang Y, Han W and Chen Z.
\newblock Effects of stratification on premixed cool flame propagation and
  modeling.
\newblock \emph{Combustion and Flame} 2021; 229: 111394.

\bibitem{Wu2018energy-fuel}
Liu F, Gao Y, Han W et~al.
\newblock The investigation on soot characteristics of gasoline/diesel blends
  in a laminar co-flow diffusion flame.
\newblock \emph{Energy \& Fuels} 2018; 32: 7841--7850.

\bibitem{Ma2018Energy}
Yao M, Ma T, Wang H et~al.
\newblock A theoretical study on the effects of thermal barrier coating on
  diesel engine combustion and emission characteristics - sciencedirect.
\newblock \emph{Energy} 2018; 162: 744--752.

\bibitem{Ma2019Analysis}
Ma T, Feng L, Wang H et~al.
\newblock Analysis of near wall combustion and pollutant migration after spray
  impingement.
\newblock \emph{International Journal of Heat and Mass Transfer} 2019; 141:
  569--579.

\bibitem{2006Direct}
Kolera-Gokula H and Echekki T.
\newblock Direct numerical simulation of premixed flame kernel-vortex
  interactions in hydrogen-air mixtures.
\newblock \emph{Combustion and Flame} 2006; 146(1/2): 155--167.

\bibitem{ni2021numerical}
Ni X, Weng C, Xu H et~al.
\newblock Numerical analysis of heat flow in wall of detonation tube during
  pulse detonation cycle.
\newblock \emph{Applied Thermal Engineering} 2021; 187: 116528.

\bibitem{watanabe2020numerical}
Watanabe H, Matsuo A, Chinnayya A et~al.
\newblock Numerical analysis of the mean structure of gaseous detonation with
  dilute water spray.
\newblock \emph{Journal of Fluid Mechanics} 2020; 887.

\bibitem{Ma2021Fuel}
Ma T, Chen D, Wang H et~al.
\newblock Influence of thermal barrier coating on partially premixed combustion
  in internal combustion engine.
\newblock \emph{Fuel} 2021; 303(2-3): 121259.

\bibitem{Wjp2012}
Zhou R and Wang JP.
\newblock Numerical investigation of flow particle paths and thermodynamic
  performance of continuously rotating detonation engines.
\newblock \emph{Combustion and Flame} 2012; 159(12): 3632--3645.

\bibitem{Xu2021AAAS}
Xu A, Shan Y, Chen F et~al.
\newblock Progress of mesoscale modeling and investigation of combustion
  multiphase flow (in chinese).
\newblock \emph{Acta Aeronautica et Astronautica Sinica} 2021; 42(12): 625842.

\bibitem{Xu2015APS}
Xu A, Zhang G and Ying Y.
\newblock Progress of discrete boltzmann modeling and simulation of combustion
  system (in chinese).
\newblock \emph{Acta Physica Sinica} 2015; 64(4): 184701.
\newblock \doi{10.7498/aps.64.184701}.

\bibitem{2016Molecular}
Liu H, Kang W, Qi Z et~al.
\newblock Molecular dynamics simulations of microscopic structure of ultra
  strong shock waves in dense helium.
\newblock \emph{Front Phys} 2016; 11(6): 197--207.

\bibitem{2017Molecular}
Liu H, Zhang Y, Kang W et~al.
\newblock Molecular dynamics simulation of strong shock waves propagating in
  dense deuterium, taking into consideration effects of excited electrons.
\newblock \emph{Phys Rev E} 2017; 95(2): 023201.

\bibitem{Xu-Chapter2}
Xu A, Zhang G and Zhang Y.
\newblock Discrete boltzmann modeling of compressible flows.
\newblock In Kyzas GZ and Mitropoulos AC (eds.) \emph{Kinetic Theory},
  chapter~02. Rijeka: InTech, 2018.
\newblock \doi{10.5772/intechopen.70748}.
\newblock \urlprefix\url{http://dx.doi.org/10.5772/intechopen.70748}.

\bibitem{Xu2021AAS}
Xu A, Chen J, Song J et~al.
\newblock Progress of discrete boltzmann study on multiphase complex flows (in
  chinese).
\newblock \emph{Acta Aerodynamica Sinica} 2021; 39(3): 138--169.

\bibitem{Xu2021CJCP}
Xu A, Song J, Chen F et~al.
\newblock Modeling and analysis methods for complex fields based on phase space
  (in chinese).
\newblock \emph{Chinese Journal of Computational Physics} published online
  2021; 38: available at
  \url{https://kns.cnki.net/kcms/detail/11.2011.O4.20210524.1535.002.html}.

\bibitem{Ji2021AIPA}
Ji Y, Lin C and Luo KH.
\newblock Three-dimensional multiple-relaxation-time discrete boltzmann model
  of compressible reactive flows with nonequilibrium effects.
\newblock \emph{AIP Advances} 2021; 11(4): 045217.

\bibitem{Lin2020Entropy}
Lin C, Su X and Zhang Y.
\newblock Hydrodynamic and thermodynamic nonequilibrium effects around shock
  waves: Based on a discrete boltzmann method.
\newblock \emph{Entropy} 2020; 22(12): 1397.

\bibitem{Lin2018CAF}
Lin C and Luo KH.
\newblock Mrt discrete boltzmann method for compressible exothermic reactive
  flows.
\newblock \emph{Computers \& Fluids} 2018; 166: 176--183.

\bibitem{Lin2018CNF}
Lin C and Luo KH.
\newblock Mesoscopic simulation of nonequilibrium detonation with discrete
  boltzmann method.
\newblock \emph{Combustion and Flame} 2018; 198: 356--362.

\bibitem{ChenL2021FOP}
Chen L, Lai H, Lin C et~al.
\newblock Specific heat ratio effects of compressible rayleigh-taylor
  instability studied by discrete boltzmann method.
\newblock \emph{Frontiers of Physics} 2021; 16(5): 52500.
\newblock \doi{10.1007/s11467-021-1096-3}.

\bibitem{Xu2018FOP}
Xu A, Zhang G, Zhang Y et~al.
\newblock Discrete boltzmann model for implosion and explosion related
  compressible flow with spherical symmetry.
\newblock \emph{Frontiers of Physics} 2018; 13(5): 135102.

\bibitem{Succi2001book}
Succi S.
\newblock In \emph{The Lattice Boltzmann Equation for fluid Dynamics and
  Beyond}. Oxford University Press, New York, 2001.

\bibitem{shan1993lattice}
Shan X and Chen H.
\newblock Lattice boltzmann model for simulating flows with multiple phases and
  components.
\newblock \emph{Physical review E} 1993; 47(3): 1815.

\bibitem{zhang2005lattice}
Zhang Y, Qin R and Emerson DR.
\newblock Lattice boltzmann simulation of rarefied gas flows in microchannels.
\newblock \emph{Physical review E} 2005; 71(4): 047702.

\bibitem{ambrus2019quadrature}
Ambru VE and Sofonea V.
\newblock \emph{Quadrature-Based Lattice Boltzmann Models for Rarefied Gas
  Flow}.
\newblock Flowing Matter, 2019.

\bibitem{chen2010multiple}
Chen F, Xu A, Zhang G et~al.
\newblock Multiple-relaxation-time lattice boltzmann approach to compressible
  flows with flexible specific-heat ratio and prandtl number.
\newblock \emph{EPL (Europhysics Letters)} 2010; 90(5): 54003.

\bibitem{li2012additional}
Li Q, Luo K, Gao Y et~al.
\newblock Additional interfacial force in lattice boltzmann models for
  incompressible multiphase flows.
\newblock \emph{Physical Review E} 2012; 85(2): 026704.

\bibitem{wang2020simple}
Wang Z, Wei Y and Qian Y.
\newblock A simple direct heating thermal immersed boundary-lattice boltzmann
  method for its application in incompressible flow.
\newblock \emph{Computers \& Mathematics with Applications} 2020; 80(6):
  1633--1649.

\bibitem{chen2018highly}
Chen Z, Shu C and Tan D.
\newblock Highly accurate simplified lattice boltzmann method.
\newblock \emph{Physics of Fluids} 2018; 30(10): 103605.

\bibitem{wang2020simplified}
Wang Y, Zhong C, Cao J et~al.
\newblock A simplified finite volume lattice boltzmann method for simulations
  of fluid flows from laminar to turbulent regime, part i: Numerical framework
  and its application to laminar flow simulation.
\newblock \emph{Computers \& Mathematics with Applications} 2020; 79(5):
  1590--1618.

\bibitem{saadat2020semi}
Saadat MH, B{\"o}sch F and Karlin IV.
\newblock Semi-lagrangian lattice boltzmann model for compressible flows on
  unstructured meshes.
\newblock \emph{Physical Review E} 2020; 101(2): 023311.

\bibitem{fei2019modeling}
Fei L, Du J, Luo KH et~al.
\newblock Modeling realistic multiphase flows using a non-orthogonal
  multiple-relaxation-time lattice boltzmann method.
\newblock \emph{Physics of Fluids} 2019; 31(4): 042105.

\bibitem{qiu2020study}
Qiu R, Bao Y, Zhou T et~al.
\newblock Study of regular reflection shock waves using a mesoscopic kinetic
  approach: Curvature pattern and effects of viscosity.
\newblock \emph{Physics of Fluids} 2020; 32(10): 106106.

\bibitem{qiu2021mesoscopic}
Qiu R, Zhou T, Bao Y et~al.
\newblock Mesoscopic kinetic approach for studying nonequilibrium hydrodynamic
  and thermodynamic effects of shock wave, contact discontinuity, and
  rarefaction wave in the unsteady shock tube.
\newblock \emph{Physical Review E} 2021; 103(5): 053113.

\bibitem{sun2020discrete}
Sun D.
\newblock A discrete kinetic scheme to model anisotropic liquid--solid phase
  transitions.
\newblock \emph{Applied Mathematics Letters} 2020; 103: 106222.

\bibitem{sun2019anisotropic}
Sun D, Xing H, Dong X et~al.
\newblock An anisotropic lattice boltzmann--phase field scheme for numerical
  simulations of dendritic growth with melt convection.
\newblock \emph{International Journal of Heat and Mass Transfer} 2019; 133:
  1240--1250.

\bibitem{zhan2021lattice}
Zhan C, Chai Z and Shi B.
\newblock A lattice boltzmann model for the coupled cross-diffusion-fluid
  system.
\newblock \emph{Applied Mathematics and Computation} 2021; 400: 126105.

\bibitem{Huang2021transition}
Huang Q, Tian F, Young J et~al.
\newblock Transition to chaos in a two-sided collapsible channel flow.
\newblock \emph{Journal of Fluid Mechanics} 2021; .

\bibitem{Wang2021Lattice}
Wang H, Tian F and Liu X.
\newblock Lattice boltzmann model for interface capturing of multiphase flows
  based on the allen-cahn equation.
\newblock \emph{Chinese Physics B} 2021; .

\bibitem{Liu2020CNF-LBM}
Liu Y, Xia J, Wan K et~al.
\newblock Simulation of char-pellet combustion and sodium release inside porous
  char using lattice boltzmann method.
\newblock \emph{Combustion and Flame} 2020; 211: 325--336.

\bibitem{2015Improvement}
Esfahanian V and Ghadyani M.
\newblock Improvement of the instability of compressible lattice boltzmann
  model by shock-detecting sensor.
\newblock \emph{Journal of Mechanical Science \& Technology} 2015; 29(5):
  1981--1991.

\bibitem{2014A}
Ghadyani M and Esfahanian V.
\newblock A more robust compressible lattice boltzmann model by using the
  numerical filters.
\newblock \emph{Journal of Mechanics} 2014; 30(05): 515--525.

\bibitem{ghadyani2015use}
Ghadyani M, Esfahanian V and Taeibi-Rahni M.
\newblock The use of shock-detecting sensor to improve the stability of lattice
  boltzmann model for high mach number flows.
\newblock \emph{International Journal of Modern Physics C} 2015; 26(01):
  1550006.

\bibitem{Xu2012PoF}
Xu AG, Zhang GC, Gan YB et~al.
\newblock Lattice boltzmann modeling and simulation of compressible flows.
\newblock \emph{Frontiers of Physics} 2012; 7(5): 582--600.

\bibitem{1995Multiple}
Xu A, Lin C, Zhang G et~al.
\newblock Multiple-relaxation-time lattice boltzmann kinetic model for
  combustion.
\newblock \emph{Physical Review E} 1995; 91(4): 043306.

\bibitem{Zhang2016CNF}
Zhang Y, Xu A, Zhang G et~al.
\newblock Kinetic modeling of detonation and effects of negative temperature
  coefficient.
\newblock \emph{Combustion and Flame} 2016; 173: 483--492.

\bibitem{Lin2016CNF}
Lin C, Xu A, Zhang G et~al.
\newblock Double-distribution-function discrete boltzmann model for combustion.
\newblock \emph{Combustion and Flame} 2016; 164: 137--151.

\bibitem{Zhang2019CPC}
Zhang Y, Xu A, Zhang G et~al.
\newblock Discrete boltzmann method for non-equilibrium flows: based on shakhov
  model.
\newblock \emph{Computer Physics Communications} 2019; 238: 50--65.

\bibitem{Zhang2018FOP}
Zhang Y, Xu A, Zhang G et~al.
\newblock Discrete ellipsoidal statistical bgk model and burnett equations.
\newblock \emph{Frontiers of Physics} 2018; 13(3): 135101.

\bibitem{Lin2021PRE}
Lin C, Luo KH, Xu A et~al.
\newblock Multiple-relaxation-time discrete boltzmann modeling of
  multicomponent mixture with nonequilibrium effects.
\newblock \emph{Physical Review E} 2021; 103(1).

\bibitem{Chen2016FOP}
Chen F, Xu A and Zhang G.
\newblock Viscosity, heat conductivity, and prandtl number effects in the
  rayleigh--taylor instability.
\newblock \emph{Frontiers of Physics} 2016; 11(6): 114703.

\bibitem{Chen2018POF}
Chen F, Xu A and Zhang G.
\newblock Collaboration and competition between richtmyer--meshkov instability
  and rayleigh--taylor instability.
\newblock \emph{Physics of Fluids} 2018; 30(10): 102105.

\bibitem{Zhang2020PoF-2Fluid}
Zhang D, Xu A, Zhang Y et~al.
\newblock Two-fluid discrete boltzmann model for compressible flows: based on
  ellipsoidal statistical bhatnagar-gross-krook.
\newblock \emph{Phys Fluids} 2020; 32: 126110.

\bibitem{Shakhov1968}
Shakhov EM.
\newblock Generalization of the krook kinetic relaxation equation.
\newblock \emph{Fluid Dynamics} 1968; 3(5): 95--96.

\end{thebibliography}

\end{document}